\documentclass[%
 amsmath,amssymb,
 aps,
 twocolumn,
 superscriptaddress,
 groupedaddress,
 prb
]{revtex4}

\usepackage{amsmath}
\usepackage{amssymb}
\usepackage{float}
\usepackage{epsfig}
\usepackage[T1]{fontenc}
\usepackage{lmodern}
\usepackage{multirow}
\usepackage{diagbox}
\usepackage{graphicx}
\usepackage{dcolumn}
\usepackage{bm}
\usepackage{hyperref}
\usepackage[mathlines]{lineno}
\usepackage{xcolor}
\usepackage{soul}
\usepackage{wrapfig}
\usepackage[english]{babel}

\def\<{\langle}
\def\>{\rangle}

\def\beq{\begin{equation}}
\def\eeq{\end{equation}}

\newcommand{\bea}{\begin{eqnarray}}
\newcommand{\eea}{\end{eqnarray}}

\def\lsim{\mathrel{\rlap{\lower4pt\hbox{\hskip1pt$\sim$}}
    \raise1pt\hbox{$<$}}}         
\def\gsim{\mathrel{\rlap{\lower4pt\hbox{\hskip1pt$\sim$}}
    \raise1pt\hbox{$>$}}}         

\usepackage{url}
\usepackage{amsfonts}
\usepackage{textcomp}
\def\BibTeX{{\rm B\kern-.05em{\sc i\kern-.025em b}\kern-.08em
    T\kern-.1667em\lower.7ex\hbox{E}\kern-.125emX}}

\begin{document}

\title{Dynamic enhancement of conductance in fractional quantum Hall constriction}

\author{Sampurna Karmakar, Amulya Ratnakar and Sourin Das \\
{\emph{Department of Physical Sciences\\
Indian Institute of Science Education and Research (IISER) Kolkata \\
Mohanpur - 741246,
West Bengal, India}}} 

\begin{abstract}
A disparity in the charge of quasi-particle excitations across a tunnel junction can trigger Andreev-like processes, creating an effect similar to that of a step-up transformer. We study such a junction in its strong coupling limit in the context of quantum Hall states. Specifically, for filling fractions $\nu=1$ and $1/3$, we show the DC gain in the transformer action is bounded by 3/2, irrespective of the interedge interaction range, while the AC gain is bounded by $\sqrt{3}$ and is sensitive to the range of the interaction. This setup presents a unique possibility of frequency-tunable resonances and anti-resonances across the QPC.
\end{abstract}                                                            
\maketitle
Proximity-induced superconducting correlation locally at the edge of a quantum Hall (QH) state could open up a possibility for the design of an almost dissipation-free mesoscopic DC voltage transformer owing to its topological immunity from backscattering. The transformer action can be understood from the fact that a positive input voltage imposed on the edge flowing into a region, where the edge is strongly coupled to a superconductor (SC), would result in a negative output voltage on the edge flowing out of that region due to Andreev process\cite{Andreev}. Such junction between QH edge and SC has been experimentally realized recently not only for integer filling fraction\cite{anindya,Amir_yacoby_2016,Rickhaus2012,Jeong,Mizuno2013,Efetov2016,Amet2016,Hatefipour2022} but also for fractional fillings\cite{Amir_recent}. However, this is expected to be a difficult route as stabilizing superconductivity at large magnetic fields is difficult. Alternatively, a similar transformer action, pointed out by Chklovskii and Halperin\cite{Halperin_Chklovskii_1998_DC_Step_up_transformer}, can also be induced by employing a junction, a QPC between two QH states such that the quasi-particle charge on the two sides of the junction is different, leading to Andreev-like process \cite{Chamon_Sandler_1998_Andreev_reflection_QH_setup}. Such an Andreev-like process was first noted by Safi and Schulz in the context of non-chiral Luttinger liquid (LL)\cite{I_Safi_1995}. A simple-minded description of this physics for the case of  $\nu=1$ and 1/3 state can be understood as follows. An incident excited quasi-particle of charge $e/3$ on the $\nu=1/3$ edge can tunnel across the junction only if it pairs up with two more quasi-particles at the junction, hence forming an electron. The tunneling process will result in two $e/3$ holes reflecting on the $\nu=1/3$ edge, similar to an Andreev reflected hole at a normal metal-superconductor junction. Consequently, such an Andreev process in the absence of superconductivity can be harvested for dissipation-less transformer action. 

Such setups based on two-dimensional electron gas (2DEG) platforms with the extended junction between $\nu=1,1/3$ has been reported ~\cite{cohen2019,Biswajit_Karmakar_2020_1_1/3_junction,Ronen2018}. A finite but not perfect Andreev reflection (accompanied by normal reflection) from a QPC between $\nu=1,1/3$ junction has also been experimentally observed in 2DEG~\cite{Hashisaka_2021_Andreev_reflection}.  More recently, a graphene-based QPC linking $\nu=1$ and $1/3$ QH edge states have been realized experimentally~\cite{Andrea_Young_transformer} and a perfect Andreev reflection limit has been achieved, paving the way for a nearly dissipationless DC voltage
step-up transformer with a gain of $3/2$. 

All the previous theoretical and experimental studies in this context primarily focused on the DC limit, which is blind to the capacitive coupling to the ambient gated environment. Hence, AC response is expected to open up new avenues both in terms of its physics and practical application of such junctions. 

\begin{figure}
\includegraphics[width=0.48\textwidth]{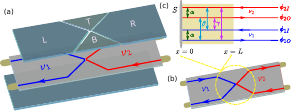}
\caption{(a) Schematic diagram motivated from Ref.[\onlinecite{Andrea_Young_transformer}] and adopted in our proposed theoretical model. In (b), the corresponding edge state configuration of a QPC has been shown schematically. In (c), $\phi_{i,I/O}$ denotes the bosonic field corresponding to the incoming/outgoing edge state of the $i^{th}$ QH edge. $\alpha,\beta,\gamma$ denote the finite range density-density repulsive interactions between the QH edges. The boundary condition at $x=0$ is denoted through the current splitting matrix $S$.}
\label{fig:bilayer-setup}
\end{figure}

An AC study comprises two complementary approaches: ({\it a}) frequency domain analysis \cite{I_Safi_1995,Safi1999,Hasisaka_2012,Hasisaka_2011_Admittance_Measurement,Plasmon,Dynamical_transport_measurement,Amit_agarwal_2014_Y_junction,Soori,HASHISAKA201832} and ({\it b}) time domain analysis\cite{I_Safi_1995,Sassetti,Soori,Amit_agarwal_2014_Y_junction,HASHISAKA201832,Kamata2014}.
We show that, in the presence of generalized interedge interactions, motivated by a recently realized experimental geometry~\cite{Andrea_Young_transformer,Cohen_four_quadrant,cohen2024} (see fig.~\ref{fig:bilayer-setup}), there exist, frequency bands, over which the AC conductance exceeds the DC bound of $3/2$. This enhancement in gain beyond the DC limit is shown to be related to the displacement current. The parameter space of interaction,  over which these bands of excess gain appear in the frequency domain, always shows distinctive behaviour in the time evolution of the electron wave packet in terms of hole pulses in the transmitted edge channel.

One of the peculiarities of AC conductance calculation for mesoscopic setup is the violation of Kirchhoff's law. This apparent current non-conservation was addressed for the non-interacting mesoscopic system by B\"uttiker et. al.~\cite{Buttiker_Dynamic_admittance,Buttiker_open_conductor,Buttiker_small_conductors} and later by Safi in the context of LLs~\cite{Safi1999}. In what follows, we will closely comply with the plasmon scattering approach developed by Safi ~\cite{Safi1997,Safi1999} for the AC conductivity matrix for such a system.

\parindent 0.1mm
\underline{\textit{Junction of Interacting Quantum Hall Edge States}}: Consider a junction of edge states realized at the boundary of two fractional QH (FQH) systems with filling fractions $\nu_{1}$ and $\nu_{2}$ in the Laughlin sequence in a four-quadrant gate geometry motivated by [\onlinecite{Andrea_Young_transformer,cohen2024}](fig.~\ref{fig:bilayer-setup}(a)). We use the corresponding effective one-dimensional model in the folded basis to describe the FQH edge junction such that all the QH edge states lie between $x=0$ and $x=\infty$, with the tunnel junction positioned at $x=0$. The distance between the edge states in the two FQH states can be adjusted by using gate electrodes, to allow for controlled interedge interactions in the region $x<L$. The two quantum wells hosting distinct FQH states can be engineered in such a way that, after the distance `$L$' from the junction, the distance between the edges becomes greater than the screening length of the interelectron interactions. As a result, the density-density interaction between the edge states gets quenched beyond $x=L$ (see fig.~\ref{fig:bilayer-setup}(b) and (c)).

We describe the low-energy physics of the FQH edge junction setup using the bosonization technique and formulate the AC conductivity matrix using the plasmon scattering matrix approach. In the standard bosonization formula, it is possible to express the fermionic field $\psi_{i,I/O}$ associated with the electron situated on $i^{th}$ edge in terms of the bosonic fields $\phi_{i,I/O}$~\cite{Haldane_1981,von_Delft_1998,maslov2005fundamental,Wen_1,Wen_2,Wen_3} as $\psi_{i,I/O} \sim F_{i,I/O} \exp(i \phi_{i,I/O}/\nu_{i})$, where $F_{I/O}$ are the corresponding Klein factors, $\nu$ denotes filling fraction and the subscript $I (O)$ describes an incoming (outgoing) bosonic field at the junction. The Lagrangian density $\mathcal{L}$ for the interacting edge, in the bosonized form, is given by
\begin{equation}
    \begin{aligned}[b]
\mathcal{L}=& -\frac{\hbar}{4\pi}\bigg(\sum_{a=1}^4\epsilon_a\partial_t\Bar{\phi}_a(x,t)\partial_x\Bar{\phi}_a(x,t)+\\
& v_F\sum_{a,b=1}^4 \int_0^\infty dx' \partial_x\Bar{\phi}_a(x,t) K_{ab}(x,x')\partial_{x'}\Bar{\phi}_b(x',t)\bigg).
\end{aligned}
\label{Eq:Lagrangian_equation}
\end{equation}

Here,  $v_F$ denotes the Fermi velocity, $(\bar{\phi}_{1},\bar{\phi}_{2},\bar{\phi}_{3},\bar{\phi}_{4})=(\frac{\phi_{1O}}{\sqrt{\nu_{1}}},\frac{\phi_{2O}}{\sqrt{\nu_{2}}},\frac{\phi_{1I}}{\sqrt{\nu_{1}}},\frac{\phi_{2I}}{\sqrt{\nu_{2}}})$, $\epsilon_a = +1 $ for $a = \{1, 2\}$, $\epsilon_a = -1 $ for $a = \{3, 4\}$ and the matrix $K(x,x')$ is given by 
\begin{widetext}
\begin{equation}
    K(x,x') =\begin{pmatrix}\delta(x-x') & V(x,x')\,\beta & -V(x,x')\,\alpha & -V(x,x')\,\gamma\\
V(x,x')\,\beta & \delta(x-x') & -V(x,x')\,\gamma & -V(x,x')\,\alpha\\
-V(x,x')\,\alpha & -V(x,x')\,\gamma & \delta(x-x') &V(x,x')\,\beta\\
-V(x,x')\,\gamma & -V(x,x')\,\alpha & V(x,x')\,\beta & \delta(x-x')\end{pmatrix}.
\label{K}
\end{equation} 
\end{widetext}
with $V(x,x')=\Theta(L-x)\Theta(L-x')\mathcal{V}(|x-x'|)$, i.e., interaction $\mathcal{V}(|x-x'|)$ is being switched on between $x=0$ and $L$. The density-density interactions are denoted as: 
$\alpha$ between $iI-iO$ edges, $\beta$ between $iI(O)-jI(O)$ edges, $\gamma$ between $iI-jO$ edges with $i,j\in\{1,2\}$ and $ i\neq j$. Various cases for the interaction $\mathcal{V}(|x-x'|)$, including interactions inspired by recent experiments, will be discussed in the subsequent section.

From Eq.~(\ref{Eq:Lagrangian_equation}), one can diagonalize the Hamiltonian and get the Heisenberg equation of motion~\cite{Sourin_Das_2009,Amulya_2021,Schulz}, and also formulate a plasmon scattering matrix, relating the oscillator modes of the incoming bosonic field to the outgoing bosonic field in the free region, by matching the fields at $x=L$ in accordance to the equation of motion (see supplemental material). The boundary condition (BC) at the tunnelling point, $x=0$, is also accounted for in the formalism. This BC is determined by the current conserving splitting matrix $\mathrm{S}$ at the junction, which may be represented by the fixed point, such that $\phi_{iO}(x=0,t) = \sum_{j=1}^{2} \mathrm{S}_{ij} \,\phi_{jI}(x=0,t)$.

The AC scattering matrix $\mathcal{S}$, relating the oscillator modes of the incoming physical fields ($\hat{c}_{iI,k} = \sqrt{\nu}_{i}\bar{c}_{iI,k}$ for $i\in \lbrace 1,2\rbrace$) to the outgoing physical modes ($\hat{c}_{iO,k} = \sqrt{\nu}_{i}\bar{c}_{iO,k}$ for $i\in \lbrace 1,2\rbrace$) in the non-interacting region ($x>L$), is given by
\begin{equation}
\begin{pmatrix}
\hat{c}_{1O,k} \\
\hat{c}_{2O,k} \\
\end{pmatrix} = \mathcal{S}\begin{pmatrix}
\hat{c}_{1I,k} \\
\hat{c}_{2I,k} \\
\end{pmatrix}
\label{Eq:AC_splitting_matrix}.
\end{equation}
There are only two allowed junction fixed points for such a setup considered here, leading to two possible current splitting matrices $\mathrm{S}$,
\begin{equation}
\mathrm{S}_{1}=\begin{pmatrix}
      1 & 0 \\
      0 & 1 
      \end{pmatrix}  \text{ and }\hspace{0.1cm}
\mathrm{S}_{2} = \frac{1}{\nu_{1}+\nu_{2}}\begin{pmatrix}
                \nu_{1} - \nu_{2} & 2\nu_{1} \\
                2\nu_{2} & \nu_{2} - \nu_{1}
                \end{pmatrix}
\label{Eq: FP}.
\end{equation}
\begin{figure*}
    \centering
    \includegraphics[width=0.98\textwidth]{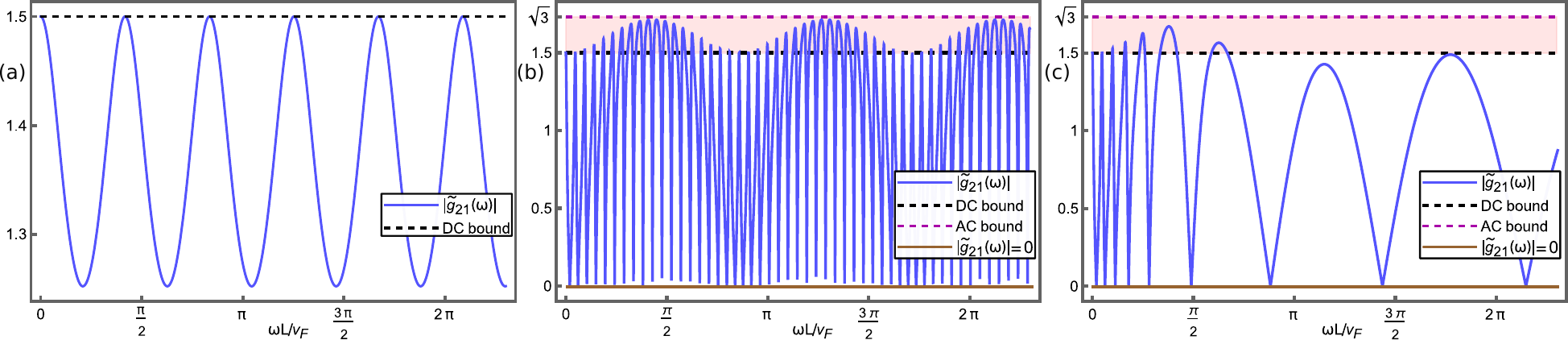}
\caption{$|\Tilde{g}_{21}(\omega)|$ is plotted as a function of frequency $\omega$ for an injection from $\nu_1=1$ side and the value for $\nu_2$ is 1/3. $|\Tilde{g}_{21}(\omega)|$ is plotted (with L = 1) for (a) $\alpha=0.55$,  (b) $\alpha=0.55,\beta=0.14 \textrm{ and }\gamma=0.589$, both with point-like interaction and (c) $\alpha=0.55,\beta=0.14,\gamma=0.589$ with screened Coulomb interaction ($\eta=12.34$). In (b) and (c), the anti-resonance points are where $|\Tilde{g}_{21}(\omega)|$ intersects the brown line.}
    \label{fig:AC_response}
\end{figure*} 
where the fully reflecting disconnected fixed point and the strongly coupled fixed point are denoted by $\mathrm{S}_{1}$ and $\mathrm{S}_{2}$~\cite{Wen_1994_junction_splitting_matrix,Line_junction,Chamon_Sandler_1998_Andreev_reflection_QH_setup} respectively.
The $2\times 2$ chiral AC conductivity matrix, $G^{S}_{AC}$, then can be written as~\cite{maslov}
\begin{equation}
    \left[G^{S}_{AC}(x,x',\omega)\right]_{ij}=(e^2/h)\nu_j\,\mathcal{S}_{ij}(\omega)\,e^{i\omega(x+x')/v_{F}},
\end{equation}
where an electron initially emanates at $x'$ and finally reaches $x$. The conductance matrix $G^{S}_{AC}$ relates the local incoming edge potential at $x'$ to the outgoing edge current at $x$, such that, $I^{S}_{O}(x,\omega) = G^{S}_{AC}(x,x',\omega)V^{S}_{I}(x',\omega)$, where $I^{S}_{O}(x,\omega) = \lbrace I_{1O},I_{2O}\rbrace^{T}_{(x,\omega)}$ and $V^{S}_{I}(x',\omega) = \lbrace V_{1I},V_{2I}\rbrace^{T}_{(x',\omega)}$. In the DC limit, that is, $\omega \rightarrow 0$, the AC scattering matrix $\mathcal{S}$ reduces to the corresponding junction fixed point matrix S (see Eq.~(\ref{Eq: FP})) at $x=0$, such that the DC conductivity matrix is given by 
\begin{equation}
    \left[G^{S}_{DC}\right]_{ij} = (e^2/h)\,\nu_{j} \left[\mathrm{S}\right]_{ij},
\end{equation}

and indicates the fact that the DC conductivity is insensitive to the density-density interactions (Eq.~(\ref{K})) present in the system.

Outgoing electron current at $x$ is related to incoming electron current at $x'$ through the relation, $I_{O}(x,\omega) = \mathcal{S}(\omega)e^{i\omega (x+x')/v_{F}}I_{I}(x',\omega)$, where $\mathcal{S}(\omega)$ has the form
\begin{eqnarray}
    \mathcal{S}(\omega) = e^{-2i\omega L/v_{F}}\begin{pmatrix}
            R_{11}(\omega) & T_{12}(\omega) \\
            T_{21} (\omega) & R_{22} (\omega)
            \end{pmatrix},
\end{eqnarray}

and $T_{ji}(\omega)(R_{ii}(\omega))$ denotes the total dynamic transmission (reflection) coefficient for the incident current from the $i^{th}$ incoming edge to the $j^{th}$ (into the $i^{th}$) outgoing edge. At the boundary of the interacting region ($x=L$),
\begin{eqnarray}
        \begin{pmatrix}
            I_{1O} \\
            I_{2O}
            \end{pmatrix}_{(L,\omega)}=
            \begin{pmatrix}
            R_{11}(\omega) & T_{12}(\omega) \\
            T_{21} (\omega) & R_{22} (\omega)
            \end{pmatrix} \begin{pmatrix}
                I_{1I} \\
                I_{2I}
            \end{pmatrix}_{(L,\omega)}.
\end{eqnarray}

It is important to note that $R_{ii}(\omega) + T_{ji}(\omega) \neq 1$ with $i,j \in \lbrace 1,2 \rbrace$ and $i\neq j$, indicating the apparent violation of the current conservation condition. Total charge conservation is restored by accounting for displacement currents, which may get induced in the capacitively coupled conductors (such as confining potential gates, QPC gates, etc.) ~\cite{Safi1999,Buttiker_small_conductors,Buttiker_Dynamic_admittance,Buttiker_open_conductor}. Assuming a one-dimensional model for capacitively coupled conductors, the net local induced current is taken to be the difference of the net local incoming current $(I_{g,I}(x,\omega))$ and a net local outgoing current $(I_{g,O}(x,\omega))$. Now, an AC current splitting matrix, $S_{AC}$, obeying the current conservation condition and relating the incoming and outgoing currents, can be defined as
$I_{O}(x) = S_{AC}\, I_{I}(x)$. $I_{O}(x)$ is given by $I_{O}(x) = \lbrace I_{1O},I_{2O},I_{g,O} \rbrace^{T}_{x}$ and $I_{I}(x)$ is given by $I_{I}(x) = \lbrace I_{1I},I_{2I},I_{g,I} \rbrace^{T}_{x}$.  An AC current conservation imposes the condition $\sum_{i=1}^{3}\left[S_{AC}\right]_{ij} = 1$. Then, a chiral $3\times 3$ AC conductance matrix, $G^{T}_{AC}$ can be formed relating the local outgoing current, $I_{O}(x,\omega)$, to the local incoming edge potential, $V_{I}(x,\omega)$, where $V_{I}(x,\omega) = \lbrace V_{1I},V_{2I},V_{g,I} \rbrace^{T}_{(x,\omega)}$ (see supplemental material). 

For a FQH edge junction of $\nu_{1}=1$ and $\nu_{2}=1/3$, tuned to strong coupling fixed point (SCFP$\left[1,\frac{1}{3}\right]$), the total $3\times 3$ current conserving conductivity matrix, $G^{T}_{AC}$ is given by
\begin{widetext}
\begin{equation}
    G^T_{AC}(L,L,\omega)= \frac{e^2}{h}\begin{pmatrix}
    R_{11}(\omega) & \frac{1}{3}T_{12}(\omega) &  1 - R_{11}(\omega) - \frac{1}{3}T_{12}(\omega) \\
    T_{21}(\omega) & \frac{1}{3}R_{22}(\omega) &  \frac{1}{3} - T_{21}(\omega) - \frac{1}{3}R_{22}(\omega) \\
    1-R_{11}(\omega)-T_{21}(\omega) & \frac{1}{3}\big(1-R_{22}(\omega)-T_{12}(\omega)\big) & R_{11}(\omega)+T_{21}(\omega) + \frac{1}{3}\big(R_{22}(\omega) + T_{12}(\omega)-1\big)
    \end{pmatrix}.  
\end{equation}
\end{widetext}
It can be easily seen, from Eq.~(\ref{Eq:AC_splitting_matrix}), that 
\begin{equation}
\mathcal{S}(\omega)N\mathcal{S}^{\dagger}(\omega) = N
\label{S_matrix},
\end{equation}

\begin{figure}
    \centering
    \includegraphics[width=0.48\textwidth]{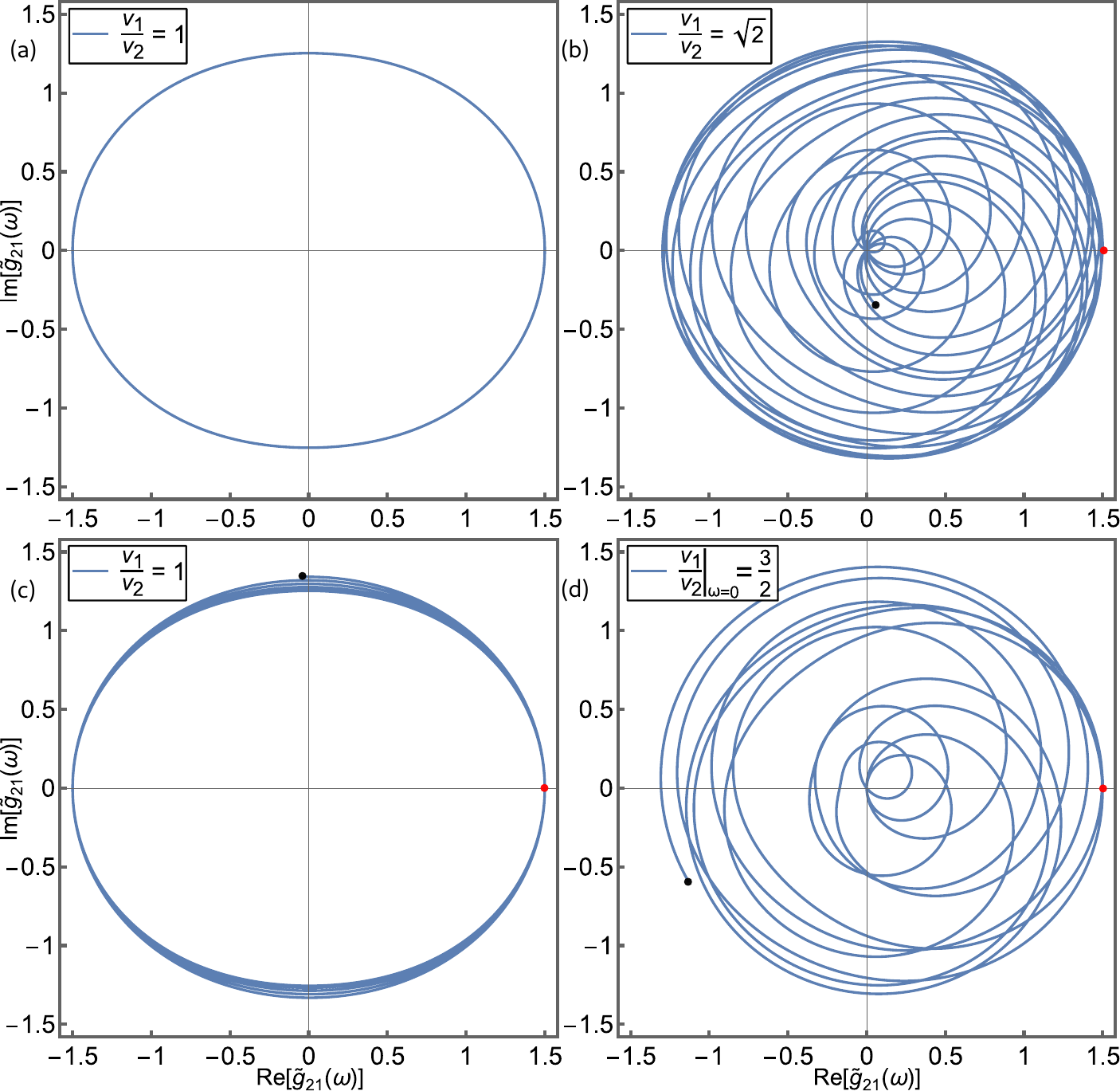}
\caption{For a QH junction of $\nu=1$ and $1/3$, $\mathrm{Im}[\Tilde{g}_{21}(\omega)]$ vs $\mathrm{Re}[\Tilde{g}_{21}(\omega)]$ is plotted as a function of frequency ($\omega$) with different $v_{1}/v_{2}$ ratios and interaction parameters for point-like (top row) and screened Coulomb ($\eta=12.34$) (bottom row) interaction. (a) and (c) are plotted for $\alpha=0.55, \beta=\gamma = 0$ ($v_{1}/v_{2}=1$). (b) is plotted for $\alpha=0.55, \beta=\gamma=0.25833$ ($v_{1}/v_{2}=\sqrt{2}$). (d) is plotted for $v_{1}/v_{2}=1.5$ at $\omega=0$ with $\alpha=0.55, \beta=\gamma=0.298077$. The red and black dots indicate the initial and final frequencies of the trajectory, respectively.}
    \label{fig:Im_Re}
\end{figure}
where $N_{ij} = \nu_{i}\,\delta_{ij}$.
Eq.~(\ref{S_matrix}) implies that for $i\neq j$, $\nu_{i}\,|R_{ii}(\omega)|^{2} + \nu_{j}\,|T_{ij}(\omega)|^{2} = \nu_{i}$. The maximum value that $|T_{ij}(\omega)|$ and $|R_{ii}(\omega)|$ can take is $\sqrt{\nu_{i}/\nu_{j}}$ and 1, respectively. Note that $|T_{12}(\omega)|/|T_{21}(\omega)|=\nu_1/\nu_2$. A voltage conversion matrix $\Tilde{G}$ with the components being $\Tilde{g}_{ij}$ can be defined as
\begin{equation}
    \begin{pmatrix}
        V_{1O}\\V_{2O}
    \end{pmatrix}_{(L,\omega)}=
       \begin{pmatrix}
        \Tilde{g}_{11}(\omega) &  \Tilde{g}_{12}(\omega)\\
         \Tilde{g}_{21}(\omega)& \Tilde{g}_{22}(\omega)
    \end{pmatrix}\begin{pmatrix}
        V_{1I}\\V_{2I}
    \end{pmatrix}_{(L,\omega)},
\end{equation}
where $\Tilde{g}_{ii}(\omega)=R_{ii}(\omega)$ and $\Tilde{g}_{ij}(\omega)=(\nu_j/\nu_i)\,T_{ij}(\omega)$ with $i \neq j$. 
For an injection from $\nu_1$ side with $\nu_1>\nu_2$, the AC gain $g_{\nu_1,\nu_2}(\omega)$, which is quantity of central interest, is defined as\cite{Andrea_Young_transformer}
\begin{equation}
   g_{\nu_1,\nu_2}(\omega)= \frac{V_{2O}(L,\omega)-V_{2I}(L,\omega)}{V_{1I}(L,\omega)-V_{2I}(L,\omega)}
\end{equation}
with $V_{2I}(L,\omega)$ being grounded and it is maximized to the value of $\Tilde{g}_{21}^{max}(\omega) = \sqrt{\nu_1/\nu_2}$. For $\nu_1=1,\nu_2=1/3,  g_{\nu_1,\nu_2}^{max}(\omega)=\sqrt{3}.$ Note that at these resonant frequencies, $|\Tilde{g}_{12/21}(\omega)|$ attains maximum value, while both $|\Tilde{g}_{11}(\omega)|$ and $|\Tilde{g}_{22}(\omega)|$ are simultaneously zero. A reverse scenario is also possible for certain frequencies for which both $|\Tilde{g}_{21}(\omega)|$ and $|\Tilde{g}_{12}(\omega)|$ are simultaneously zero (anti-resonance points), while $|\Tilde{g}_{11/22}(\omega)|$ attains the maximum value of one. 

\underline{\textit{Choice of $\mathcal{V}(|x-x'|)$}}:  (1) First, we investigate a point-like interaction, i.e., $\mathcal{V}(|x-x'|)=\delta(x-x')$,  to establish a basic framework for our investigation. The $\Tilde{G}$ matrix can be formulated by following the plasmon scattering approach discussed in the previous section. 

For a junction of $\nu_{1} = 1$ and $\nu_{2}=1/3$, which already shows an enhanced transmitted current in the DC limit $(|\Tilde{g}_{21}(\omega = 0)| = 3/2)$, in the presence of only intraedge interaction  ($\beta,\gamma =0$), the AC current amplitude does not go beyond its DC limit, i.e., $|\Tilde{g}_{21}(\omega)| \leq 3/2$ (as shown in fig.~\ref{fig:AC_response}(a)). However, in the presence of interedge interaction ($\beta, \gamma \neq 0$), when the current is excited along the $\nu = 1$ incoming edge, the transmitted current amplitude may even go beyond its DC limit, resulting in a maximum gain, $g_{\nu_1=1,\nu_2=1/3}(\omega)= \sqrt 3$, at resonant frequencies (see fig.~\ref{fig:AC_response}(b)). At these magic frequencies, the reflectance
vanishes. The excess current beyond unit amplitude is compensated by the displacement current in the capacitive-coupled ambient gate. Hence, this provides a basic setup to realize a step-up AC transformer with higher efficiency as compared to its DC counterpart in a QPC geometry.\\
(2) For the long-range interaction, the screened Coulomb potential, $\mathcal{V}(|x-x'|)=\frac{e^{-\eta |x-x'|}}{|x-x'|}$ (where $1/\eta$ represents the range of interaction), characterizes the interaction among all electrons within a finite length $L$.  This formulation is primarily motivated by Ref.[\onlinecite{Cohen_four_quadrant,cohen2024}] and the $\eta$ parameter, which is fitted from these experiments, has a value of 12.34 (see supplemental material). The profile for $\Tilde{g}_{21}(\omega)$ shows an enhancement beyond the DC limit for lower frequencies (see fig.~\ref{fig:AC_response}(c)).

In fig.~\ref{fig:Im_Re}, we study the non-trivial dependence of conductance on frequency, where a parametric plot of $\mathrm{Im}[\Tilde{g}_{21}(\omega)]$, $\mathrm{Re}[\Tilde{g}_{21}(\omega)]$ is presented as function of frequency ($\omega$) for different $v_{1}/v_{2}$ ratios. Note that with point-like interaction and  $v_{1}/v_{2} = 1$, i.e., for $\alpha \neq 0$ and $\beta= \gamma = 0$, there is only one timescale corresponding to the time of flight of plasmons in the cavity, hence $\mathrm{Im}[\Tilde{g}_{21}(\omega)]$ vs $\mathrm{Re}[\Tilde{g}_{21}(\omega)]$ is a simple close loop (see fig.~\ref{fig:Im_Re}(a)) for $\nu_{1}=1,\nu_{2}=1/3$. However, for irrational ratios of $v_{1}/v_{2}$ (see fig.~\ref{fig:Im_Re}(b)), where for simplicity, we have taken $\beta = \gamma$, we note that the loop does not close onto itself, indicating a lack of periodicity in the system within the frequency domain, and interestingly, fig.~\ref{fig:Im_Re}(b) also depicts transmission zeros (anti-resonances). For the case of long-range interactions, the system also exhibits no periodicity in the frequency domain, regardless of whether $v_1/v_2$ is rational or irrational (see fig.~\ref{fig:Im_Re}(c) and (d)).

The above conductances can also be obtained via a wave packet dynamics approach \cite{I_Safi_1995,Sassetti,Soori,Amit_agarwal_2014_Y_junction,HASHISAKA201832,Kamata2014} in the time domain (see supplemental material). Interestingly, we note that in response to an injected wave packet, a distinct nature of fractionalized negative pulses in the transmitted edge channel appears, which does not appear for intraedge point-like interactions.

\underline{\textit{Conclusion:}} 
We present a study of SCFP$\left[\nu_1,\nu_2\right]$ in presence of finite range interaction in the AC regime and show that an appropriate set-up design could open up a new avenue in terms of potential applications in quantum-inspired technologies such as step-up transformers with amplification beyond the experimentally observed DC limit and give rise to resonances and anti-resonances at magic frequencies, which we can make use of as frequency filters. As $\eta$ increases, the finite range screened Coulomb interaction approaches the point-like interaction limit, and additional resonance peaks emerge within the same frequency window (see section VI of supplemental material). Furthermore, for different ratios of $v_1$ and $v_2$, the periodicity of the conductance as a function of frequency in the complex plane is studied in fig.~\ref{fig:Im_Re}, which has been discussed experimentally in Ref.[\onlinecite{Feve}]. We also note that a realistic model of QPC-induced interaction region of length of few $\mathrm{\mu}$m~\cite{Feve,Moty,Bauerle_2018,Kataoka,Freulon2015} with the parameter space of interaction strength that we are working with, the estimated range for resonance frequency can be of the order of 1-15 GHz for typical propagation velocity of $10^4\sim 10^5$  m/s~\cite{Bauerle_2018,Freulon2015,Kataoka,sim,Safi_magnetoplasmon} for the plasmon, which is well within experimental reach~\cite{high_frequency1,high_frequency2,highfrequency3,Feve}.  Hence, the results presented in this letter are significant in light of recent experiments.

\underline{\textit{Acknowledgements:}}
The authors would like to thank Biswajit Karmakar and Suvankar Purkait for their useful discussion regarding AC transport in QH circuits and, and Taige Wang and Andrea Young for their email conversations. A.R. acknowledges the University Grants Commission, India, for support through fellowship. S.K. is thankful to the Ministry of Education of the Government of India for financial support through the Prime Minister's Research Fellows (PMRF) grant.
\bibliography{reference}

@article{Andreev,
title = {THERMAL CONDUCTIVITY OF THE INTERMEDIATE STATE OF SUPERCONDUCTORS},
author = {Andreev, A F},
abstractNote = {},
doi = {},
url = {https://www.osti.gov/biblio/4071988}, journal = {Zh. Eksperim. i Teor. Fiz.},
place = {Country unknown/Code not available},
year = {1964},
month = {5}
}

@article{anindya,
  title = {Inter-Landau-level Andreev Reflection at the Dirac Point in a Graphene Quantum Hall State Coupled to a ${\mathrm{NbSe}}_{2}$ Superconductor},
  author = {Sahu, Manas Ranjan and Liu, Xin and Paul, Arup Kumar and Das, Sourin and Raychaudhuri, Pratap and Jain, J. K. and Das, Anindya},
  journal = {Phys. Rev. Lett.},
  volume = {121},
  issue = {8},
  pages = {086809},
  numpages = {6},
  year = {2018},
  month = {Aug},
  publisher = {American Physical Society},
  doi = {10.1103/PhysRevLett.121.086809},
  url = {https://link.aps.org/doi/10.1103/PhysRevLett.121.086809}
}

@Article{Amir_yacoby_2016,
author={Lee, Gil-Ho
and Huang, Ko-Fan
and Efetov, Dmitri K.
and Wei, Di S.
and Hart, Sean
and Taniguchi, Takashi
and Watanabe, Kenji
and Yacoby, Amir
and Kim, Philip},
title={Inducing superconducting correlation in quantum Hall edge states},
journal={Nature Physics},
year={2017},
month={Jul},
day={01},
volume={13},
number={7},
pages={693-698},
issn={1745-2481},
doi={10.1038/nphys4084},
url={https://doi.org/10.1038/nphys4084}
}

@Article{Rickhaus2012,
author={Rickhaus, Peter
and Weiss, Markus
and Marot, Laurent
and Sch{\"o}nenberger, Christian},
title={Quantum Hall Effect in Graphene with Superconducting Electrodes},
journal={Nano Letters},
year={2012},
month={Apr},
day={11},
publisher={American Chemical Society},
volume={12},
number={4},
pages={1942-1945},
issn={1530-6984},
doi={10.1021/nl204415s},
url={https://doi.org/10.1021/nl204415s}
}

@article{Jeong,
  title = {Observation of supercurrent in PbIn-graphene-PbIn Josephson junction},
  author = {Jeong, Dongchan and Choi, Jae-Hyun and Lee, Gil-Ho and Jo, Sanghyun and Doh, Yong-Joo and Lee, Hu-Jong},
  journal = {Phys. Rev. B},
  volume = {83},
  issue = {9},
  pages = {094503},
  numpages = {5},
  year = {2011},
  month = {Mar},
  publisher = {American Physical Society},
  doi = {10.1103/PhysRevB.83.094503},
  url = {https://link.aps.org/doi/10.1103/PhysRevB.83.094503}
}

@Article{Mizuno2013,
author={Mizuno, Naomi
and Nielsen, Bent
and Du, Xu},
title={Ballistic-like supercurrent in suspended graphene Josephson weak links},
journal={Nature Communications},
year={2013},
month={Nov},
day={14},
volume={4},
number={1},
pages={2716},
issn={2041-1723},
doi={10.1038/ncomms3716},
url={https://doi.org/10.1038/ncomms3716}
}

@Article{Hatefipour2022,
author={Hatefipour, Mehdi
and Cuozzo, Joseph J.
and Kanter, Jesse
and Strickland, William M.
and Allemang, Christopher R.
and Lu, Tzu-Ming
and Rossi, Enrico
and Shabani, Javad},
title={Induced Superconducting Pairing in Integer Quantum Hall Edge States},
journal={Nano Letters},
year={2022},
month={Aug},
day={10},
publisher={American Chemical Society},
volume={22},
number={15},
pages={6173-6178},
issn={1530-6984},
doi={10.1021/acs.nanolett.2c01413},
url={https://doi.org/10.1021/acs.nanolett.2c01413}
}

@Article{Efetov2016,
author={Efetov, D. K.
and Wang, L.
and Handschin, C.
and Efetov, K. B.
and Shuang, J.
and Cava, R.
and Taniguchi, T.
and Watanabe, K.
and Hone, J.
and Dean, C. R.
and Kim, P.},
title={Specular interband Andreev reflections at van der Waals interfaces between graphene and NbSe2},
journal={Nature Physics},
year={2016},
month={Apr},
day={01},
volume={12},
number={4},
pages={328-332},
issn={1745-2481},
doi={10.1038/nphys3583},
url={https://doi.org/10.1038/nphys3583}
}

@article{
Amet2016,
author = {F. Amet  and C. T. Ke  and I. V. Borzenets  and J. Wang  and K. Watanabe  and T. Taniguchi  and R. S. Deacon  and M. Yamamoto  and Y. Bomze  and S. Tarucha  and G. Finkelstein },
title = {Supercurrent in the quantum Hall regime},
journal = {Science},
volume = {352},
number = {6288},
pages = {966-969},
year = {2016},
doi = {10.1126/science.aad6203},
URL = {https://www.science.org/doi/abs/10.1126/science.aad6203}}

@article{Amir_recent,
  title = {Andreev Reflection in the Fractional Quantum Hall State},
  author = {G\"ul, \"Onder and Ronen, Yuval and Lee, Si Young and Shapourian, Hassan and Zauberman, Jonathan and Lee, Young Hee and Watanabe, Kenji and Taniguchi, Takashi and Vishwanath, Ashvin and Yacoby, Amir and Kim, Philip},
  journal = {Phys. Rev. X},
  volume = {12},
  issue = {2},
  pages = {021057},
  numpages = {23},
  year = {2022},
  month = {Jun},
  publisher = {American Physical Society},
  doi = {10.1103/PhysRevX.12.021057},
  url = {https://link.aps.org/doi/10.1103/PhysRevX.12.021057}
}

@article{Halperin_Chklovskii_1998_DC_Step_up_transformer,
  title = {Consequences of a possible adiabatic transition between $\ensuremath{\nu}=1/3$ and $\ensuremath{\nu}=1$ quantum Hall states in a narrow wire},
  author = {Chklovskii, Dmitri B. and Halperin, Bertrand I.},
  journal = {Phys. Rev. B},
  volume = {57},
  issue = {7},
  pages = {3781--3784},
  numpages = {0},
  year = {1998},
  month = {Feb},
  publisher = {American Physical Society},
  doi = {10.1103/PhysRevB.57.3781},
  url = {https://link.aps.org/doi/10.1103/PhysRevB.57.3781}
}

@article{Chamon_Sandler_1998_Andreev_reflection_QH_setup,
  title = {Andreev reflection in the fractional quantum Hall effect},
  author = {Sandler, Nancy P. and Chamon, Claudio de C. and Fradkin, Eduardo},
  journal = {Phys. Rev. B},
  volume = {57},
  issue = {19},
  pages = {12324--12332},
  numpages = {0},
  year = {1998},
  month = {May},
  publisher = {American Physical Society},
  doi = {10.1103/PhysRevB.57.12324},
  url = {https://link.aps.org/doi/10.1103/PhysRevB.57.12324}
}

@article{I_Safi_1995,
  title = {Transport in an inhomogeneous interacting one-dimensional system},
  author = {Safi, I. and Schulz, H. J.},
  journal = {Phys. Rev. B},
  volume = {52},
  issue = {24},
  pages = {R17040--R17043},
  numpages = {0},
  year = {1995},
  month = {Dec},
  publisher = {American Physical Society},
  doi = {10.1103/PhysRevB.52.R17040},
  url = {https://link.aps.org/doi/10.1103/PhysRevB.52.R17040}
}

@article{Biswajit_Karmakar_2020_1_1/3_junction,
  title = {Magnetic-Field-Dependent Equilibration of Fractional Quantum Hall Edge Modes},
  author = {Maiti, Tanmay and Agarwal, Pooja and Purkait, Suvankar and Sreejith, G. J. and Das, Sourin and Biasiol, Giorgio and Sorba, Lucia and Karmakar, Biswajit},
  journal = {Phys. Rev. Lett.},
  volume = {125},
  issue = {7},
  pages = {076802},
  numpages = {6},
  year = {2020},
  month = {Aug},
  publisher = {American Physical Society},
  doi = {10.1103/PhysRevLett.125.076802},
  url = {https://link.aps.org/doi/10.1103/PhysRevLett.125.076802}
}

@Article{Ronen2018,
author={Ronen, Yuval
and Cohen, Yonatan
and Banitt, Daniel
and Heiblum, Moty
and Umansky, Vladimir},
title={Robust integer and fractional helical modes in the quantum Hall effect},
journal={Nature Physics},
year={2018},
month={Apr},
day={01},
volume={14},
number={4},
pages={411-416},
issn={1745-2481},
doi={10.1038/s41567-017-0035-2},
url={https://doi.org/10.1038/s41567-017-0035-2}
}

@Article{cohen2019,
author={Cohen, Yonatan
and Ronen, Yuval
and Yang, Wenmin
and Banitt, Daniel
and Park, Jinhong
and Heiblum, Moty
and Mirlin, Alexander D.
and Gefen, Yuval
and Umansky, Vladimir},
title={Synthesizing a $\nu$=2/3 fractional quantum Hall effect edge state from counter-propagating $\nu$=1 and $\nu$=1/3 states},
journal={Nature Communications},
year={2019},
month={Apr},
day={23},
volume={10},
number={1},
pages={1920},
issn={2041-1723},
doi={10.1038/s41467-019-09920-5},
url={https://doi.org/10.1038/s41467-019-09920-5}
}

@Article{Hashisaka_2021_Andreev_reflection,
author={Hashisaka, M.
and Jonckheere, T.
and Akiho, T.
and Sasaki, S.
and Rech, J.
and Martin, T.
and Muraki, K.},
title={Andreev reflection of fractional quantum Hall quasiparticles},
journal={Nature Communications},
year={2021},
month={May},
day={14},
volume={12},
number={1},
pages={2794},
issn={2041-1723},
doi={10.1038/s41467-021-23160-6},
url={https://doi.org/10.1038/s41467-021-23160-6}
}

@article{
Andrea_Young_transformer,
author = {Liam A. Cohen  and Noah L. Samuelson  and Taige Wang  and Takashi Taniguchi  and Kenji Watanabe  and Michael P. Zaletel  and Andrea F. Young },
title = {Universal chiral Luttinger liquid behavior in a graphene fractional quantum Hall point contact},
journal = {Science},
volume = {382},
number = {6670},
pages = {542-547},
year = {2023},
doi = {10.1126/science.adf9728},
URL = {https://www.science.org/doi/abs/10.1126/science.adf9728}}

@Article{Safi1999,
author={Safi, I.},
title={A dynamic scattering approach for a gated interacting wire},
journal={The European Physical Journal B - Condensed Matter and Complex Systems},
year={1999},
month={Dec},
day={01},
volume={12},
number={3},
pages={451-455},
issn={1434-6036},
doi={10.1007/s100510051026},
url={https://doi.org/10.1007/s100510051026}
}

@article{Amit_agarwal_2014_Y_junction,
  title = {Time-resolved transport properties of a Y junction of Tomonaga-Luttinger liquid wires},
  author = {Agarwal, Amit},
  journal = {Phys. Rev. B},
  volume = {90},
  issue = {19},
  pages = {195403},
  numpages = {11},
  year = {2014},
  month = {Nov},
  publisher = {American Physical Society},
  doi = {10.1103/PhysRevB.90.195403},
  url = {https://link.aps.org/doi/10.1103/PhysRevB.90.195403}
}

@article{Soori,
doi = {10.1209/0295-5075/93/57007},
url = {https://dx.doi.org/10.1209/0295-5075/93/57007},
year = {2011},
month = {mar},
publisher = {},
volume = {93},
number = {5},
pages = {57007},
author = {Abhiram Soori and Diptiman Sen},
title = {Conductance of Tomonaga-Luttinger liquid wires and junctions with resistances},
journal = {Europhysics Letters}
}

@article{Hasisaka_2012,
  title = {Distributed electrochemical capacitance evidenced in high-frequency admittance measurements on a quantum Hall device},
  author = {Hashisaka, Masayuki and Washio, Kazuhisa and Kamata, Hiroshi and Muraki, Koji and Fujisawa, Toshimasa},
  journal = {Phys. Rev. B},
  volume = {85},
  issue = {15},
  pages = {155424},
  numpages = {5},
  year = {2012},
  month = {Apr},
  publisher = {American Physical Society},
  doi = {10.1103/PhysRevB.85.155424},
  url = {https://link.aps.org/doi/10.1103/PhysRevB.85.155424}
}

@article{Hasisaka_2011_Admittance_Measurement,
doi = {10.1143/JJAP.50.04DJ04},
url = {https://dx.doi.org/10.1143/JJAP.50.04DJ04},
year = {2011},
month = {apr},
publisher = {},
volume = {50},
number = {4S},
pages = {04DJ04},
author = {Kazuhisa Washio and Masayuki Hashisaka and Hiroshi Kamata and Koji Muraki and Toshimasa Fujisawa},
title = {Admittance Measurement for a Quantum Point Contact in a Multiterminal Quantum Hall Device},
journal = {Japanese Journal of Applied Physics}
}

@article{HASHISAKA201832,
title = {Tomonaga–Luttinger-liquid nature of edge excitations in integer quantum Hall edge channels},
journal = {Reviews in Physics},
volume = {3},
pages = {32-43},
year = {2018},
issn = {2405-4283},
doi = {https://doi.org/10.1016/j.revip.2018.07.001},
url = {https://www.sciencedirect.com/science/article/pii/S2405428318300078},
author = {Masayuki Hashisaka and Toshimasa Fujisawa}
}

@Article{Kamata2014,
author={Kamata, H.
and Kumada, N.
and Hashisaka, M.
and Muraki, K.
and Fujisawa, T.},
title={Fractionalized wave packets from an artificial Tomonaga--Luttinger liquid},
journal={Nature Nanotechnology},
year={2014},
month={Mar},
day={01},
volume={9},
number={3},
pages={177-181},
issn={1748-3395},
doi={10.1038/nnano.2013.312},
url={https://doi.org/10.1038/nnano.2013.312}
}

@article{Plasmon,
  title = {Plasmon scattering approach to energy exchange and high-frequency noise in $\ensuremath{\nu}=2$ quantum Hall edge channels},
  author = {Degiovanni, P. and Grenier, Ch. and F\`eve, G. and Altimiras, C. and le Sueur, H. and Pierre, F.},
  journal = {Phys. Rev. B},
  volume = {81},
  issue = {12},
  pages = {121302},
  numpages = {4},
  year = {2010},
  month = {Mar},
  publisher = {American Physical Society},
  doi = {10.1103/PhysRevB.81.121302},
  url = {https://link.aps.org/doi/10.1103/PhysRevB.81.121302}
}

@article{Dynamical_transport_measurement,
  title = {Dynamical transport measurement of the Luttinger parameter in helical edges states of two-dimensional topological insulators},
  author = {M\"uller, Tobias and Thomale, Ronny and Trauzettel, Bj\"orn and Bocquillon, Erwann and Kashuba, Oleksiy},
  journal = {Phys. Rev. B},
  volume = {95},
  issue = {24},
  pages = {245114},
  numpages = {7},
  year = {2017},
  month = {Jun},
  publisher = {American Physical Society},
  doi = {10.1103/PhysRevB.95.245114},
  url = {https://link.aps.org/doi/10.1103/PhysRevB.95.245114}
}

@article{Sassetti,
  title = {Time-resolved pure spin fractionalization and spin-charge separation in helical Luttinger liquid based devices},
  author = {Calzona, Alessio and Carrega, Matteo and Dolcetto, Giacomo and Sassetti, Maura},
  journal = {Phys. Rev. B},
  volume = {92},
  issue = {19},
  pages = {195414},
  numpages = {8},
  year = {2015},
  month = {Nov},
  publisher = {American Physical Society},
  doi = {10.1103/PhysRevB.92.195414},
  url = {https://link.aps.org/doi/10.1103/PhysRevB.92.195414}
}

@article{Buttiker_Dynamic_admittance,
  title = {Dynamic admittance of mesoscopic conductors: Discrete-potential model},
  author = {Pr\^etre, A. and Thomas, H. and B\"uttiker, M.},
  journal = {Phys. Rev. B},
  volume = {54},
  issue = {11},
  pages = {8130--8143},
  numpages = {0},
  year = {1996},
  month = {Sep},
  publisher = {American Physical Society},
  doi = {10.1103/PhysRevB.54.8130},
  url = {https://link.aps.org/doi/10.1103/PhysRevB.54.8130}
}

@article{Buttiker_small_conductors,
  title = {Dynamic conductance and the scattering matrix of small conductors},
  author = {B\"uttiker, M. and Pr\^etre, A. and Thomas, H.},
  journal = {Phys. Rev. Lett.},
  volume = {70},
  issue = {26},
  pages = {4114--4117},
  numpages = {0},
  year = {1993},
  month = {Jun},
  publisher = {American Physical Society},
  doi = {10.1103/PhysRevLett.70.4114},
  url = {https://link.aps.org/doi/10.1103/PhysRevLett.70.4114}
}

@article{Buttiker_open_conductor ,
  title = {Scattering theory of thermal and excess noise in open conductors},
  author = {B\"uttiker, M.},
  journal = {Phys. Rev. Lett.},
  volume = {65},
  issue = {23},
  pages = {2901--2904},
  numpages = {0},
  year = {1990},
  month = {Dec},
  publisher = {American Physical Society},
  doi = {10.1103/PhysRevLett.65.2901},
  url = {https://link.aps.org/doi/10.1103/PhysRevLett.65.2901}
}

@article{Safi1997,
  title = {Conductance of a quantum wire: Landauer's approach versus the Kubo formula},
  author = {Safi, In`es},
  journal = {Phys. Rev. B},
  volume = {55},
  issue = {12},
  pages = {R7331--R7334},
  numpages = {0},
  year = {1997},
  month = {Mar},
  publisher = {American Physical Society},
  doi = {10.1103/PhysRevB.55.R7331},
  url = {https://link.aps.org/doi/10.1103/PhysRevB.55.R7331}
}

@article{Amulya_2021,
  title = {Enhancement in tunneling density of states in a Luttinger liquid: Role of nonlocal interaction},
  author = {Ratnakar, Amulya and Das, Sourin},
  journal = {Phys. Rev. B},
  volume = {104},
  issue = {4},
  pages = {045402},
  numpages = {13},
  year = {2021},
  month = {Jul},
  publisher = {American Physical Society},
  doi = {10.1103/PhysRevB.104.045402},
  url = {https://link.aps.org/doi/10.1103/PhysRevB.104.045402}
}

@article{Haldane_1981,
	doi = {10.1088/0022-3719/14/19/010},
	url = {https://doi.org/10.1088/0022-3719/14/19/010},
	year = 1981,
	month = {jul},
	publisher = {{IOP} Publishing},
	volume = {14},
	number = {19},
	pages = {2585--2609},
	author = {F. D. M. Haldane},
	    title = "{Luttinger liquid theory of one-dimensional quantum fluids. I. Properties of the Luttinger model and their extension to the general 1D interacting spinless Fermi gas}",
}

@article{von_Delft_1998,
	doi = {10.1002/andp.19985100401},
  
	url = {https://doi.org/10.1002%2Fandp.19985100401},
  
	year = 1998,
	month = {nov},
  
	publisher = {Wiley},
  
	volume = {510},
  
	number = {4},
  
	pages = {225--305},
  
	author = {Jan von Delft and Herbert Schoeller},
  
	title = {Bosonization for beginners {\textemdash} refermionization for experts},
  
	journal = {Annalen der Physik}
}

@misc{maslov2005fundamental,
      title={Fundamental aspects of electron correlations and quantum transport in one-dimensional systems}, 
      author={Dmitrii L. Maslov},
      year={2005},
      eprint={cond-mat/0506035},
      archivePrefix={arXiv},
      primaryClass={cond-mat.mes-hall},
url={https://doi.org/10.48550/arXiv.cond-mat/0506035}
}

@article{Wen_1,
  title = {Chiral Luttinger liquid and the edge excitations in the fractional quantum Hall states},
  author = {Wen, X. G.},
  journal = {Phys. Rev. B},
  volume = {41},
  issue = {18},
  pages = {12838--12844},
  numpages = {0},
  year = {1990},
  month = {Jun},
  publisher = {American Physical Society},
  doi = {10.1103/PhysRevB.41.12838},
  url = {https://link.aps.org/doi/10.1103/PhysRevB.41.12838}
}

@article{Wen_2,
  title = {Electrodynamical properties of gapless edge excitations in the fractional quantum Hall states},
  author = {Wen, X. G.},
  journal = {Phys. Rev. Lett.},
  volume = {64},
  issue = {18},
  pages = {2206--2209},
  numpages = {0},
  year = {1990},
  month = {Apr},
  publisher = {American Physical Society},
  doi = {10.1103/PhysRevLett.64.2206},
  url = {https://link.aps.org/doi/10.1103/PhysRevLett.64.2206}
}

@article{Wen_3,
  title = {Edge transport properties of the fractional quantum Hall states and weak-impurity scattering of a one-dimensional charge-density wave},
  author = {Wen, Xiao-Gang},
  journal = {Phys. Rev. B},
  volume = {44},
  issue = {11},
  pages = {5708--5719},
  numpages = {0},
  year = {1991},
  month = {Sep},
  publisher = {American Physical Society},
  doi = {10.1103/PhysRevB.44.5708},
  url = {https://link.aps.org/doi/10.1103/PhysRevB.44.5708}
}

@article{Sourin_Das_2009,
doi = {10.1209/0295-5075/86/37010},
url = {https://dx.doi.org/10.1209/0295-5075/86/37010},
year = {2009},
month = {may},
publisher = {},
volume = {86},
number = {3},
pages = {37010},
author = {Sourin Das and Sumathi Rao and Diptiman Sen},
title = {Effect of inter-edge Coulomb interactions on transport through a point contact in a ν= 5/2 quantum Hall state},
journal = {Europhysics Letters}
}

@article{Wen_1994_junction_splitting_matrix,
  title = {Impurity effects on chiral one-dimensional electron systems},
  author = {Wen, Xiao-Gang},
  journal = {Phys. Rev. B},
  volume = {50},
  issue = {8},
  pages = {5420--5428},
  numpages = {0},
  year = {1994},
  month = {Aug},
  publisher = {American Physical Society},
  doi = {10.1103/PhysRevB.50.5420},
  url = {https://link.aps.org/doi/10.1103/PhysRevB.50.5420}
}

@article{Line_junction,
  title = {Line junction in a quantum Hall system with two filling fractions},
  author = {Sen, Diptiman and Agarwal, Amit},
  journal = {Phys. Rev. B},
  volume = {78},
  issue = {8},
  pages = {085430},
  numpages = {9},
  year = {2008},
  month = {Aug},
  publisher = {American Physical Society},
  doi = {10.1103/PhysRevB.78.085430},
  url = {https://link.aps.org/doi/10.1103/PhysRevB.78.085430}
}

@article{maslov,
  title = {Transport through dirty Luttinger liquids connected to reservoirs},
  author = {Maslov, Dmitrii L.},
  journal = {Phys. Rev. B},
  volume = {52},
  issue = {20},
  pages = {R14368--R14371},
  numpages = {0},
  year = {1995},
  month = {Nov},
  publisher = {American Physical Society},
  doi = {10.1103/PhysRevB.52.R14368},
  url = {https://link.aps.org/doi/10.1103/PhysRevB.52.R14368}
}

@Article{Feve,
author={Bocquillon, E.
and Freulon, V.
and Berroir, J.-.. M.
and Degiovanni, P.
and Pla{\c{c}}ais, B.
and Cavanna, A.
and Jin, Y.
and F{\`e}ve, G.},
title={Separation of neutral and charge modes in one-dimensional chiral edge channels},
journal={Nature Communications},
year={2013},
month={May},
day={14},
volume={4},
number={1},
pages={1839},
issn={2041-1723},
doi={10.1038/ncomms2788},
url={https://doi.org/10.1038/ncomms2788}
}

@article{Bauerle_2018,
doi = {10.1088/1361-6633/aaa98a},
url = {https://dx.doi.org/10.1088/1361-6633/aaa98a},
year = {2018},
month = {apr},
publisher = {IOP Publishing},
volume = {81},
number = {5},
pages = {056503},
author = {Christopher Bäuerle and D Christian Glattli and Tristan Meunier and Fabien Portier and Patrice Roche and Preden Roulleau and Shintaro Takada and Xavier Waintal},
title = {Coherent control of single electrons: a review of current progress},
journal = {Reports on Progress in Physics}
}

@article{Kataoka,
  title = {Time-of-Flight Measurements of Single-Electron Wave Packets in Quantum Hall Edge States},
  author = {Kataoka, M. and Johnson, N. and Emary, C. and See, P. and Griffiths, J. P. and Jones, G. A. C. and Farrer, I. and Ritchie, D. A. and Pepper, M. and Janssen, T. J. B. M.},
  journal = {Phys. Rev. Lett.},
  volume = {116},
  issue = {12},
  pages = {126803},
  numpages = {5},
  year = {2016},
  month = {Mar},
  publisher = {American Physical Society},
  doi = {10.1103/PhysRevLett.116.126803},
  url = {https://link.aps.org/doi/10.1103/PhysRevLett.116.126803}
}

@article{Moty,
  title = {Charge Fractionalization in the Integer Quantum Hall Effect},
  author = {Inoue, Hiroyuki and Grivnin, Anna and Ofek, Nissim and Neder, Izhar and Heiblum, Moty and Umansky, Vladimir and Mahalu, Diana},
  journal = {Phys. Rev. Lett.},
  volume = {112},
  issue = {16},
  pages = {166801},
  numpages = {5},
  year = {2014},
  month = {Apr},
  publisher = {American Physical Society},
  doi = {10.1103/PhysRevLett.112.166801},
  url = {https://link.aps.org/doi/10.1103/PhysRevLett.112.166801}
}

@Article{Freulon2015,
author={Freulon, V.
and Marguerite, A.
and Berroir, J.-M.
and Pla{\c{c}}ais, B.
and Cavanna, A.
and Jin, Y.
and F{\`e}ve, G.},
title={Hong-Ou-Mandel experiment for temporal investigation of single-electron fractionalization},
journal={Nature Communications},
year={2015},
month={Apr},
day={21},
volume={6},
number={1},
pages={6854},
issn={2041-1723},
doi={10.1038/ncomms7854},
url={https://doi.org/10.1038/ncomms7854}
}

@article{sim,
  title = {Partition of Two Interacting Electrons by a Potential Barrier},
  author = {Ryu, Sungguen and Sim, H.-S.},
  journal = {Phys. Rev. Lett.},
  volume = {129},
  issue = {16},
  pages = {166801},
  numpages = {7},
  year = {2022},
  month = {Oct},
  publisher = {American Physical Society},
  doi = {10.1103/PhysRevLett.129.166801},
  url = {https://link.aps.org/doi/10.1103/PhysRevLett.129.166801}
}

@Article{Cohen_four_quadrant,
author={Cohen, Liam A.
and Samuelson, Noah L.
and Wang, Taige
and Klocke, Kai
and Reeves, Cian C.
and Taniguchi, Takashi
and Watanabe, Kenji
and Vijay, Sagar
and Zaletel, Michael P.
and Young, Andrea F.},
title={Nanoscale electrostatic control in ultraclean van der Waals heterostructures by local anodic oxidation of graphite gates},
journal={Nature Physics},
year={2023},
month={Oct},
day={01},
volume={19},
number={10},
pages={1502-1508},
issn={1745-2481},
doi={10.1038/s41567-023-02114-3},
url={https://doi.org/10.1038/s41567-023-02114-3}
}

@misc{cohen2024,
      author={Liam A. Cohen and Noah L. Samuelson and Taige Wang and Kai Klocke and Cian C. Reeves and Takashi Taniguchi and Kenji Watanabe and Sagar Vijay and Michael P. Zaletel and Andrea F. Young},
      year={2024},
      eprint={2401.10433},
      archivePrefix={arXiv},
      primaryClass={cond-mat.mes-hall},
      url={https://doi.org/10.48550/arXiv.2401.10433}
}

@misc{Safi_magnetoplasmon,
      author={E. Frigerio and G. Rebora and M. Ruelle and H. Souquet-Basiège and Y. Jin and U. Gennser and A. Cavanna and B. Plaçais and E. Baudin and J. -M. Berroir and I. Safi and P. Degiovanni and G. Fève and G. Ménard},
      year={2024},
      eprint={2404.18204},
      archivePrefix={arXiv},
      primaryClass={cond-mat.mes-hall},
url={https://doi.org/10.48550/arXiv.2404.18204}
}

@article{high_frequency1,
  title = {Microwave Hall conductivity of the two-dimensional electron gas in GaAs-${\mathrm{Al}}_{\mathrm{x}}$${\mathrm{Ga}}_{1\mathrm{\ensuremath{-}}\mathrm{x}}$As},
  author = {Kuchar, F. and Meisels, R. and Weimann, G. and Schlapp, W.},
  journal = {Phys. Rev. B},
  volume = {33},
  issue = {4},
  pages = {2965--2967},
  numpages = {0},
  year = {1986},
  month = {Feb},
  publisher = {American Physical Society},
  doi = {10.1103/PhysRevB.33.2965},
  url = {https://link.aps.org/doi/10.1103/PhysRevB.33.2965}
}

@article{high_frequency2,
  title = {Dynamical Scaling of the Quantum Hall Plateau Transition},
  author = {Hohls, F. and Zeitler, U. and Haug, R. J. and Meisels, R. and Dybko, K. and Kuchar, F.},
  journal = {Phys. Rev. Lett.},
  volume = {89},
  issue = {27},
  pages = {276801},
  numpages = {4},
  year = {2002},
  month = {Dec},
  publisher = {American Physical Society},
  doi = {10.1103/PhysRevLett.89.276801},
  url = {https://link.aps.org/doi/10.1103/PhysRevLett.89.276801}
}

@article{highfrequency3,
title = {Quantum Hall effect from finite-frequency studies},
journal = {Physica B: Condensed Matter},
volume = {227},
number = {1},
pages = {173-179},
year = {1996},
note = {Proceedings of the Third International Symposium on New Phenomena in Mesoscopic Structures},
issn = {0921-4526},
doi = {https://doi.org/10.1016/0921-4526(96)00391-2},
url = {https://www.sciencedirect.com/science/article/pii/0921452696003912},
author = {L.W. Engel and Y.P. Li and D.C. Tsui}
}

@article{Schulz,
  title = {Wigner crystal in one dimension},
  author = {Schulz, H. J.},
  journal = {Phys. Rev. Lett.},
  volume = {71},
  issue = {12},
  pages = {1864--1867},
  numpages = {0},
  year = {1993},
  month = {Sep},
  publisher = {American Physical Society},
  doi = {10.1103/PhysRevLett.71.1864},
  url = {https://link.aps.org/doi/10.1103/PhysRevLett.71.1864}
}
\newpage
\widetext{
\begin{center}
\textbf{\large Supplemental material for $``$Dynamic enhancement of conductance in fractional quantum Hall constriction$"$}\\

Sampurna Karmakar, Amulya Ratnakar and Sourin Das \\
Department of Physical Sciences\\
Indian Institute of Science Education and Research (IISER) Kolkata \\
Mohanpur - 741246,
West Bengal, India
\end{center}
\numberwithin{equation}{section}
\section{Brief description of the geometry} Fig. 1(a) in the main text shows an upper gate with four segments: top (T), bottom (B), left (L), and right (R). The L and R gates adjust filling factors by fixing electron density, while the T and B gates create a QPC constriction by depleting the electron gas below them. In our study, we have adopted the corresponding effective one-dimensional model based on a folded geometry.
\section{Formulation of AC conductance matrix}
Consider Fig.~1 of the main text. The interacting edge Hamiltonian describing our setup, in bosonized form, is given by
\begin{equation}
    H = \mathcal{H}+H_g,
\end{equation}
where the kinetic and the interacting part of the total Hamiltonian $(H)$ is given by $\mathcal{H}$, and the coupling to the ambient gates is accounted by $H_g$.
\begin{eqnarray}
     \mathcal{H} &=& \frac{\hbar v_F}{4\pi} \sum_{a,b=1}^4\int_0^\infty dx  \int_0^\infty dx' \bigg(\partial_x\Bar{\phi}_a(x,t) K_{ab}(x,x')\partial_{x'}\Bar{\phi}_b(x',t)\bigg),\nonumber\\
H_g&=& eV_g\int_{0}^\infty  dx\sum_{i=1}^2\,(\rho_{iI}(x)+\rho_{iO}(x))= eV_g \,Q,
\end{eqnarray}
where the incoming/outgoing electronic density,  $\rho_{iI/O}(x)=\mp(1/2\pi)\,\partial_x\phi_{iI/O}(x)$ for $i\in\lbrace 1,2\rbrace$, $v_F$ is the Fermi velocity, $(\bar{\phi}_{1},\bar{\phi}_{2},\bar{\phi}_{3},\bar{\phi}_{4})=(\frac{\phi_{1O}}{\sqrt{\nu_{1}}},\frac{\phi_{2O}}{\sqrt{\nu_{2}}},\frac{\phi_{1I}}{\sqrt{\nu_{1}}},\frac{\phi_{2I}}{\sqrt{\nu_{2}}})$ and the matrix $K(x,x')$ is given by 
\begin{equation}
        K(x,x') =\begin{pmatrix}\delta(x-x') & V(x,x')\,\beta & -V(x,x')\,\alpha & -V(x,x')\,\gamma\\
V(x,x')\,\beta & \delta(x-x') & -V(x,x')\,\gamma & -V(x,x')\,\alpha\\
-V(x,x')\,\alpha & -V(x,x')\,\gamma & \delta(x-x') &V(x,x')\,\beta\\
-V(x,x')\,\gamma & -V(x,x')\,\alpha & V(x,x')\,\beta & \delta(x-x')\end{pmatrix}.
\label{kmat}
\end{equation}
with $V(x,x')=\Theta(L-x)\Theta(L-x')\mathcal{V}(|x-x'|)$, i.e., an interaction $\mathcal{V}(|x-x'|)$ is being switched on between $x=0$ and $L$. We will closely comply with the plasmon scattering approach developed by I. Safi~\cite{Safi1999} for the calculation of the AC conductivity matrix of our system.

A gauge transformation of the chiral fermionic fields yields a substitution for the chiral bosonic fields and the corresponding densities as
\begin{eqnarray}
    \phi_{iI/O}(x) &\rightarrow&   \phi_{iI/O}(x)+\frac{e}{\hbar}\int^x A(x')dx',\\
    \rho_{iI/O}(x) &\rightarrow&  \rho_{iI/O}(x)\mp\frac{e}{h}A(x),
\end{eqnarray}
and the presence of a vector potential `$A$' modifies the Hamiltonian as
\begin{equation}
     H(\rho_{1/2 I},\rho_{1/2 O})\rightarrow H^{(A)}=H\left(\rho_{1/2 I}-\frac{eA}{h}\,,\,\rho_{1/2 O}+
     \frac{eA}{h}\right) .
\end{equation}
From here, the current field $I(x)$ can be expressed for the Hamiltonian $H^{(A)}$ as
\begin{equation}
     I(x) = -\frac{\delta H^{(A)}}{\delta A}\Big|_{A=0}.
\end{equation}
Hence, $I(x)$ takes the form
\begin{equation}
    I_{iI/O}(x)=\pm\frac{e}{h}\mu_{iI/O}(x).
\end{equation}
where, $\mu_{iI/O}(x)=\frac{\delta H}{\delta \rho_{iI/O}}$ is the local chemical potential and the local potential $V_{iI/O}(x)$ is given by $V_{iI/O}(x) = \mu_{iI/O}(x)/e$. 
Using these relations, one can formulate the AC current splitting matrix $S_{AC}$ relating all the incoming currents at a point $x$ to the outgoing currents at the same point $x$ as
\begin{eqnarray}
    \begin{pmatrix}
    \mathrm{I}_{1O}(x,\omega) \\ \mathrm{I}_{2O}(x,\omega) \\\mathrm{I}_{gO}(x,\omega)
\end{pmatrix} &=& \begin{pmatrix}
    R_{11}(\omega) &  T_{12}(\omega) & T_{1g}(\omega)\\ T_{21}(\omega) & R_{22}(\omega) & T_{2g}(\omega)\\ T_{g1}(\omega) & T_{g2}(\omega) & R_{gg}(\omega)
\end{pmatrix}\begin{pmatrix}
    \mathrm{I}_{1I}(x,\omega) \\ \mathrm{I}_{2I}(x,\omega)\\\mathrm{I}_{gI}(x,\omega)
\end{pmatrix}\nonumber\\ &=& S_{AC}(\omega)\begin{pmatrix}
    \mathrm{I}_{1I}(x,\omega) \\ \mathrm{I}_{2I}(x,\omega)\\\mathrm{I}_{gI}(x,\omega)
\end{pmatrix},
\end{eqnarray}
where $T_{ji}(R_{ii})$ denotes the total dynamic transmission(reflection) coefficient for the incident current from the $i^{th}$ incoming edge to the $j^{th}$ outgoing edge (into the $i^{th}$ outgoing edge). The requirement of AC current conservation imposes the condition that the columns of the matrix $S_{AC}(\omega)$ should add up to 1, i.e.
\begin{eqnarray}
R_{11}(\omega)+T_{21}(\omega) + T_{g1}(\omega)&=&1,\nonumber\\
T_{12}(\omega)+ R_{22}(\omega) + T_{g2}(\omega)&=&1,\nonumber\\
T_{1g}(\omega)+T_{2g}(\omega)+R_{gg}(\omega)&=&1.
\label{current_conservation}
\end{eqnarray}
Again, the relation between voltages and currents on the $i^{th}$ edges is, $I_i(x,\omega)=\frac{e^2}{h}\nu_i\,V_i(x,\omega)$. Hence,
\begin{equation}
    \begin{pmatrix}
    \mathrm{V}_{1O}(x,\omega) \\ \mathrm{V}_{2O}(x,\omega) \\\mathrm{V}_{gO}(x,\omega)
\end{pmatrix} =M^{-1}S_{AC}(\omega)\,M\begin{pmatrix}
    \mathrm{V}_{1I}(x,\omega) \\ \mathrm{V}_{2I}(x,\omega)\\\mathrm{V}_{gI}(x,\omega)
\end{pmatrix}.
\end{equation} where the elements of the matrix $M$ are given by the filling fraction of the two quantum Hall (QH) bulk and the ambient gates, such that $M=\text{diag}\,(1,\frac{1}{3},1)$.
If all the edges have the same input bias voltage, each row of the matrix $[M^{-1}S_{AC}(\omega)\,M]$ should add up to 1, such that the voltages at the outgoing edges are also the same. This leads to the following conditions
\begin{eqnarray}
    R_{11}(\omega) + \frac{1}{3} T_{12}(\omega) + T_{1g}(\omega) &=& 1,\nonumber\\
     3T_{21}(\omega) + R_{22}(\omega) + 3T_{2g}(\omega) &=& 1,\nonumber\\
      T_{g1}(\omega) + \frac{1}{3} T_{g2}(\omega) + R_{gg}(\omega) &=& 1
      \label{voltage_bias}.
\end{eqnarray}
Now, the conductance matrix calculated at the interaction region boundary, $x=L$, is given by $G^T_{AC}(L, L,\omega)$ as
    \begin{equation}
       \begin{pmatrix}
    \mathrm{I}_{1O}(L,\omega) \\ \mathrm{I}_{2O}(L,\omega) \\\mathrm{I}_{gO}(L,\omega)
\end{pmatrix} = G^T_{AC}(L, L,\omega) \begin{pmatrix}
    \mathrm{V}_{1I}(L,\omega) \\ \mathrm{V}_{2I}(L,\omega) \\\mathrm{V}_{gI}(L,\omega)
\end{pmatrix} .
\end{equation}
Using (\ref{current_conservation}) and (\ref{voltage_bias}), and the relation
$G^T_{AC}(L, L,\omega)=\frac{e^2}{h}S_{AC}(\omega)M$
\begin{equation}
    G^T_{AC}(L,L,\omega)= \frac{e^2}{h}\begin{pmatrix}
    R_{11}(\omega) & \frac{1}{3}T_{12}(\omega) &  1 - R_{11}(\omega) - \frac{1}{3}T_{12}(\omega) \\
    T_{21}(\omega) & \frac{1}{3}R_{22}(\omega) &  
    \frac{1}{3}- T_{21}(\omega) - \frac{1}{3}R_{22}(\omega) \\
    1-R_{11}(\omega)-T_{21}(\omega) & \frac{1}{3}\big(1-R_{22}(\omega)-T_{12}(\omega)\big) & R_{11}(\omega)+T_{21}(\omega) + \frac{1}{3}\big(R_{22}(\omega) + T_{12}(\omega)\big) -\frac{1}{3} 
    \end{pmatrix}.  
\end{equation}
$G^{S}_{AC}$, which is the $2\times 2$ part of the conductance matrix $G^{T}_{AC}(L,L,\omega)$, can be calculated using the plasmon scattering matrix approach~\cite{I_Safi_1995,Sourin_Das_2009,Amulya_2021}.
\section{Conductance matrix calculation}
The Lagrangian density, $\mathcal{L}$, only for the interacting edges in the bosonized form, is given by
\begin{equation}
\mathcal{L}= -\frac{\hbar}{4\pi}\bigg(\sum_{a=1}^4\epsilon_a\partial_t\Bar{\phi}_a(x,t)\partial_x\Bar{\phi}_a(x,t)+ v_F\sum_{a,b=1}^4 \int_0^\infty dx' \partial_x\Bar{\phi}_a(x,t) K_{ab}(x,x')\partial_{x'}\Bar{\phi}_b(x',t)\bigg).
\label{Eq:LagrangianL}
\end{equation}
 Here, $\epsilon_a = +1 $ for $a = \{1, 2\}$, $\epsilon_a = -1 $ for $a = \{3, 4\}$ and $K_{ab}(x,x')$ is the matrix accounting for coordinate-dependent interedge interaction, which has been mentioned in Eq.~(\ref{kmat}). The commutation relation between the physical bosonic fields is given by $\left[\phi_{i,I/O}(x),\phi_{j,I/O}(y)\right] = \mp i\pi \,\nu_{i}\,\delta_{ij}\,\mathrm{sgn}(x-y)$. One can get the equation of motions from Eq.~(\ref{Eq:LagrangianL}) as
\begin{eqnarray}
    \partial_x\left(\epsilon_a \partial_t\Bar{\phi}_{a}(x,t)+ v_F\sum_b\int_0^\infty dx' K_{ab}(x,x')\partial_{x'}\Bar{\phi}_{b}(x',t)\right) &=0&
    \label{Eq:Eq_of_motionL}.
\end{eqnarray}

The boundary condition (BC) at $x=0$ is determined by the current conserving splitting matrix $\mathrm{S}$ at the junction, such that
\begin{equation}
    \phi_{iO}(x=0,t) = \sum_{j=1}^{2} \mathrm{S}_{ij}\, \phi_{jI}(x=0,t).
\end{equation}
The BC at $x=L$ can be obtained by integrating Eq.~(\ref{Eq:Eq_of_motionL}) over the coordinate $x$ from $L-\epsilon$ to $L+\epsilon$, which leads to the condition $\phi_{iO/I}(x=L-\epsilon,t)=\phi_{iO/I}(x=L+\epsilon,t)$ in the limit $\epsilon \rightarrow 0$, i.e., the physical bosonic fields should be continuous at the interacting region boundary $x=L$. 

At time $t=0$, the physical bosonic field can be expressed in mode expansion form as
\begin{equation}
\bar{\phi}_{a}(x) = \int_{0}^{\infty} \frac{dk}{k} \left(\bar{c}_{ak}e^{i\epsilon_{a}kx} + \bar{c}^{\dagger}_{ak}e^{-i\epsilon_{a}kx}\right),   
\end{equation}
where, $\bar{c}_{\alpha,k}$ ($\bar{c}^{\dagger}_{\alpha,k}$) is the bosonic annihilation (creation) operator, such that $\left[\bar{c}_{\alpha,k}, \bar{c}^{\dagger}_{\alpha',k'}\right] = \delta_{\alpha \alpha'}k\,\delta(k-k')$. The commutation relation for the physical fields is given by $\left[\bar{\phi}_{a}(x),\bar{\phi}_{a'}(y)\right] =  i\pi\,\epsilon_{a}\, \delta_{aa'} \,\mathrm{sgn}(x-y)$. To diagonalize the Lagrangian in Eq.~(\ref{Eq:LagrangianL}) using Bogoliubov (Bg) transformation, let us now define Bg transformed fields 
\begin{equation}
    \tilde{\phi}_{\alpha}(x,t)=\int_0^\infty \frac{dk}{k}\, \left(\Tilde{c}_{\alpha k}e^{i \epsilon
    _\alpha k (x-v_\alpha t)}+\Tilde{c}_{\alpha k}^\dagger e^{-i \epsilon
    _\alpha k (x-v_\alpha t)}\right),
\end{equation}
where $\alpha\in \lbrace 1,2,3,4\rbrace$ , $v_{\alpha}$ is the renormalized velocity, $\epsilon_\alpha=$ sgn$(v_\alpha)$ and $\tilde{c}_{\alpha,k}$ ($\tilde{c}^{\dagger}_{\alpha,k}$) is the bosonic annihilation (creation) operator for the $\alpha$-th bosonic mode, such that $\left[\tilde{c}_{\alpha k},\tilde{c}^{\dagger}_{\alpha'k'}\right] = \delta_{\alpha\alpha'}k\,\delta(k-k')$. Inside the interacting region, that is, $0<x<L$, the physical interacting fields can be related to the chiral Bg oscillator modes through a real matrix $X(k)$ as 
\begin{equation}
\bar{\phi}_{a}(x,t)=\sum_\alpha\int_{0}^{\infty} \frac{dk}{k}\, X_{a \alpha}(k)  \left(\tilde{c}_{\alpha k}e^{ik\epsilon_{\alpha}\left(x-v_{\alpha}(k)t\right)} + \bar{c}^{\dagger}_{ak}e^{-ik\epsilon_{\alpha}(x-v_{\alpha}(k)t)}\right).
\label{bogoliubov}
\end{equation}
The interacting bosonic oscillator modes $\bar{c}_{a k}$  can be related to the Bg oscillator modes $\tilde{c}_{\alpha k}$ as
\begin{equation}
    \bar{c}_{ak}=\sum_{\alpha=1}^4 X_{a\alpha}(k)\left(P_{a\alpha,+}\tilde{c}_{\alpha k}e^{i(\epsilon_\alpha-\epsilon_a)kx}+P_{a\alpha,-}\tilde{c}_{\alpha k}e^{-i(\epsilon_\alpha+\epsilon_a)kx}\right),
\end{equation}
where the projection operator $P_{a\alpha,\pm}$ is given by 
\begin{equation}
    P_{a\alpha,\pm}=\frac{1}{2}(1\pm\epsilon_a \epsilon_\alpha ).
\end{equation}
To obtain the $X(k)$ matrix, let's write the Heisenberg equation of motion
\begin{equation}
   \frac{d}{dt}\Bar{\phi}_a(x,t)=-v_F\epsilon_{a}\int_0^\infty dx'\sum_{\alpha = 1}^{4} K_{a\alpha}(x,x')\frac{d}{dx'}\bar{\phi}_{\alpha}(x',t).
   \label{heisenberg}
\end{equation}
Using Eq.~(\ref{bogoliubov}) and ~(\ref{heisenberg}),  we get an eigenvalue equation to solve for $X(k)$ and $v_\alpha(k)$, which is given by
\begin{equation}
    v_F\sum_{b=1}^4 \epsilon_a K_{ab}(\epsilon_\alpha k) X_{b\alpha}(k)=X_{a\alpha}(k)v_\alpha(k).
\end{equation}
The interelectron interaction between the FQH edge states renormalizes the Fermi velocity such that 
\begin{eqnarray}
    v_{1}(k) &=& v_{F}\sqrt{(1+\beta\mathbb{V}(k))^{2} - (\alpha\mathbb{V}(k) + \gamma\mathbb{V}(k))^{2}},\\ \nonumber
    v_{2}(k) &=& v_{F}\sqrt{(1-\beta\mathbb{V}(k))^{2} - (\alpha\mathbb{V}(k)-\gamma\mathbb{V}(k))^{2}} ,
\end{eqnarray}
where $\mathbb{V}(k)$ is the Fourier transformed form for the interaction term $\mathcal{V}(|x-x'|)$. Now let's write the matrix $X(k)$ as 
\begin{equation}
    X(k)=\begin{pmatrix}
X_1(k) & X_2(k)\\ X_3(k) & X_4(k)\end{pmatrix}. 
\end{equation}
If we write $\Tilde{c}_{Ok}=\begin{pmatrix}
    \Tilde{c}_{1Ok}  &  \Tilde{c}_{2Ok}
\end{pmatrix}^T$, $\Tilde{c}_{Ik}=\begin{pmatrix}
    \Tilde{c}_{1Ik}  &  \Tilde{c}_{2Ik}
\end{pmatrix}^T$, then taking into account the BC at $x=0$ will lead to the relation 
\begin{equation}
    \Tilde{c}_{Ok}=\Tilde{S}(k)  \Tilde{c}_{Ik},
\end{equation}
where  
\begin{equation}
    \Tilde{S}(k) = \left(X_{1}(k) - \bar{\mathrm{S}}X_{3}(k)\right)^{-1}\left(\bar{\mathrm{S}}X_{4}(k)-X_{2}(k)\right),
\end{equation}and 
\begin{equation}
    \bar{\mathrm{S}} = M^{-1}\mathrm{S}M,
\end{equation}  
with $\left[M\right]_{ij} = \sqrt{\nu_{i}}\,\delta_{ij}$.

Since the system is non-dissipative, we impose the condition, $k_{F}v_{F} = kv_{i}(k)$ and that the current spitting matrix $\mathrm{S}$ at the junction is given by the current conserving fixed point of the theory. The continuity of bosonic fields $\bar{\phi}_{a}$ at $x=L$ ensures that the current is conserved across the interacting region boundary for individual right/left moving edge states. Then, the matrix relating the oscillator modes of the incoming physical fields ($\hat{c}_{i,I,k} = \sqrt{\nu}_{i}\bar{c}_{j,k}$ for $j=i+2$ with $i\in \lbrace 1,2\rbrace$) to the outgoing physical modes ($\hat{c}_{i,k} = \sqrt{\nu}_{i}\bar{c}_{i,O,k}$ for $i\in \lbrace 1,2\rbrace$) in the non-interacting region ($x>L$), is given by
\begin{eqnarray}
\begin{pmatrix}
\hat{c}_{1O,k} \\
\hat{c}_{2O,k} \\
\end{pmatrix} = \mathcal{S}\begin{pmatrix}
\hat{c}_{1I,k} \\
\hat{c}_{2I,k} \\
\end{pmatrix}
\label{Eq:Current_splitting_matrix},
\end{eqnarray}
where, $\mathcal{S}=e^{-2ikL}\xi\,\Gamma^{-1}$, such that
\begin{eqnarray}
\xi &=& X_{1}(k)W(k,L)\Tilde{S}(k) + X_{2}(k)W(k,L)^{-1}, \nonumber\\
\Gamma &=& X_{3}(k)W(k,L)\Tilde{S}(k) + X_{4}(k)W(k,L)^{-1},
\end{eqnarray}
and
\begin{equation}
W(k,L) = \begin{pmatrix}
e^{ikLv_F/v_1(k)} & 0 \\
0 & e^{ikLv_F/v_2(k)}
\end{pmatrix}.
\end{equation}
The BC at $x=0$ may be presented by the fixed point of the theory, and the only two allowed junction fixed points for such a setup considered here are given by
\begin{equation}
\mathrm{S}_{1}=\begin{pmatrix}
      1 & 0 \\
      0 & 1 
      \end{pmatrix}  \text{ and }\hspace{0.1cm}
\mathrm{S}_{2} = \frac{1}{\nu_{1}+\nu_{2}}\begin{pmatrix}
                \nu_{1} - \nu_{2} & 2\nu_{1} \\
                2\nu_{2} & \nu_{2} - \nu_{1}
                \end{pmatrix},
\end{equation}
where the fully reflecting disconnected fixed point and the strongly coupled fixed point are denoted by $\mathrm{S}_{1}$ and $\mathrm{S}_{2}$~\cite{Wen_1994_junction_splitting_matrix,Line_junction,Chamon_Sandler_1998_Andreev_reflection_QH_setup} respectively.
\section{Relation between conductance matrix and the $\mathcal{S}$ matrix }
At temperature $T=0$, $\langle \hat{c}_{mI,k}\, \hat{c}_{jI,k'}\rangle=0=\langle \hat{c}_{mI,k}^\dagger \, \hat{c}_{jI,k'}^\dagger \rangle \text{ and }\langle \hat{c}_{mI,k} \, \hat{c}_{jI,k'}^\dagger\rangle=\nu_j\,\delta_{mj} \,k\,\delta(k-k')$ as $\langle \hat{c}_{jI,k}^\dagger \, \hat{c}_{jI,k}\rangle=0$.
The $2\times 2$ chiral AC conductivity matrix, $G^{S}_{AC}$ can be expressed in terms of the current-current correlation function using the Kubo formula as 
\begin{equation}
    \left[G^{S}_{AC}(x,x',\omega)\right]_{ij}=\frac{e^2\omega}{h}\mathcal{G}_{ij}(x,x',\omega),
\end{equation}
where $\mathcal{G}_{ij}(x,x',\omega)$ is given by 
\begin{equation}
\mathcal{G}_{ij}(x,x',\omega)= \int_0^{\infty}\, dt\,\Big\langle T_t\,\phi_{iO}(x,t)\,\phi_{jI}(x',0)\Big\rangle \,e^{i\omega t}.
\end{equation}
Note that both $x$ and $x'$ are in non-interacting region ($x>L$). To perform the integral, the physical bosonic fields can be expanded in the corresponding oscillator modes such that
\begin{align}
\mathcal{G}_{ij}(x,x',\omega) &= \int_0^{\infty}dt\int_0^{\infty}\frac{dk}{k}\int_0^{\infty}\frac{dk'}{k'}\Bigg\langle\bigg(\sum_m \Big(\mathcal{S}_{im}\,\hat{c}_{mI,k}\,e^{ik(x-v_Ft)}+ \hat{c}_{mI,k}^\dagger\, \mathcal{S}_{mi}^*\,e^{-ik(x-v_Ft)}\Big)\bigg)\bigg( \hat{c}_{jI,k'}\,e^{-ik'x'}\nonumber\\
& \hspace{1cm}+ \hat{c}_{jI,k'}^\dagger \,e^{ik'x'}\bigg)\Bigg\rangle \,e^{i\omega t}\nonumber\\
&=  \int_0^{\infty}dt\int_0^{\infty}\frac{dk}{k}\int_0^{\infty}\frac{dk'}{k'}\sum_m \mathcal{S}_{im}\langle \hat{c}_{mI,k}\, \hat{c}^\dagger_{jI,k'}\rangle\, e^{ikx+ik'x'}\,e^{-ikv_Ft}\,e^{i\omega t}\nonumber\\
&=  \int_0^{\infty}dt\int_0^{\infty}\frac{dk}{k}\int_0^{\infty}\frac{dk'}{k'}\sum_m \mathcal{S}_{im}\,\nu_j\,\delta_{mj}\,k\,\delta(k-k') 
   \,e^{ikx+ik'x'}\, e^{-ikv_Ft} \,e^{i\omega t}\nonumber\\
&=  \int_0^{\infty}dt\int_0^{\infty}\frac{dk}{k}\,\mathcal{S}_{ij}\,\nu_j\,e^{ik(x+x')}\, e^{-ikv_Ft} \,e^{i\omega t}\nonumber\\
&=\int_0^{\infty}\frac{dk}{k}\,\mathcal{S}_{ij}\,\nu_j\,e^{ik(x+x')}\,\delta(\omega-kv_F)\label{Eq:non-dissipative}\\
&=  \frac{1}{\omega}\nu_j\,\mathcal{S}_{ij}(\omega)\,e^{i\omega(x+x')/v_F}
\end{align}
Hence 
\begin{equation}
\left[G^{S}_{AC}(x,x',\omega)\right]_{ij}=\frac{e^2}{h}\nu_j\,\mathcal{S}_{ij}(\omega)\,e^{i\omega(x+x')/v_F}.    
\end{equation}
Note that, for a non-dissipative system, the frequency of the external electric field $\omega$ matches with the system's natural frequency, which is equal to $kv_F$ (see Eq.~(\ref{Eq:non-dissipative})).
\section{Choice of long-range interaction} 
\begin{figure}[h]
    \centering
    \includegraphics[width=0.5\linewidth]{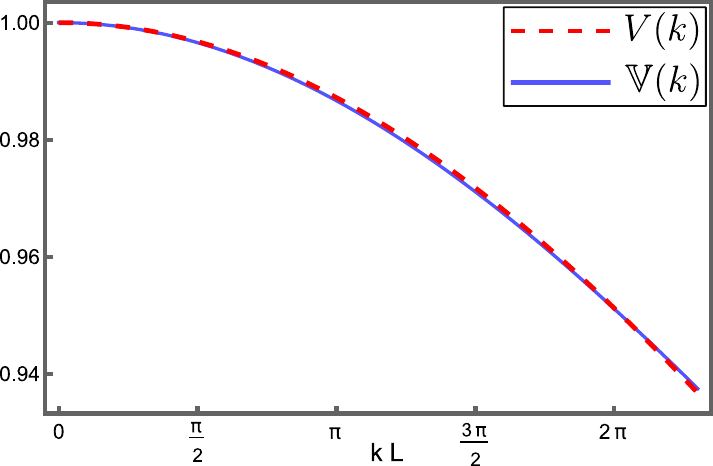}
   \caption{$V(k)$ and $\mathbb{V}(k)$ have been plotted together with the parameters $d_t=30$ nm and $d_b=15$ nm for $V(k)$ and $d=0.01$ and $\eta=12.34$ for $\mathbb{V}(k)$.}
\label{fig:fit}
\end{figure}
For the choice of long-range interaction, we have chosen the screened Coulomb potential, $\mathcal{V}(|x-x'|)=\frac{e^{-\eta |x-x'|}}{|x-x'|}$ and denoted the corresponding Fourier transformed term as $\mathbb{V}(k)$. Note that, while performing the Fourier transformation, we avoid the singularity at the origin by using the approximate form of the interaction as $\frac{e^{-\eta|x-x'|}}{\sqrt{(x-x')^2+d^2}}$. This choice fits very closely to an interaction as given in Ref.~[\onlinecite{Cohen_four_quadrant,cohen2024}],
\begin{equation}
    V(k)=\frac{e^2}{4\pi\epsilon_0 \epsilon_{\textrm{hBN}}}\frac{4\pi\sinh{(B d_t|k|)} \sinh{(B d_b|k|
)}}{\sinh{(B (d_t+d_b)|k|)|k|}},
\end{equation}
where $d_t$ and $d_b$ are the distances of the upper four gates and the lower gate, respectively, from the middle layer, which is filled by hBN with dielectric constant $\epsilon_\perp = 3$ and $\epsilon_\parallel = 6.6, B=\sqrt{\epsilon_\parallel/\epsilon_\perp},\epsilon_{\textrm{hBN}}=\sqrt{\epsilon_\parallel \epsilon_\perp}$. 
In Fig.~\ref{fig:fit}, we have treated all the length scales of the problem by rescaling them with respect to the interacting region length of 500nm. With such a choice of parameters, it can be shown that (see Fig.~\ref{fig:fit}) $\mathbb{V}(k)$ fits quite well with $V(k)$ after normalizing both of the functions to be equal 1 at the origin. Hence, it is very significant for leading to the realization of an experimentally relevant system based on our model. Note that when the gates come closer to the middle layer, the screening effect due to the presence of adjacent gates becomes more effective, thereby reducing the overall prominence of the long-range effect.
\section{Finite frequency conductance for different screening lengths of Coulomb interaction}
\begin{figure}[h]
    \centering
    \includegraphics[width=0.98\textwidth]{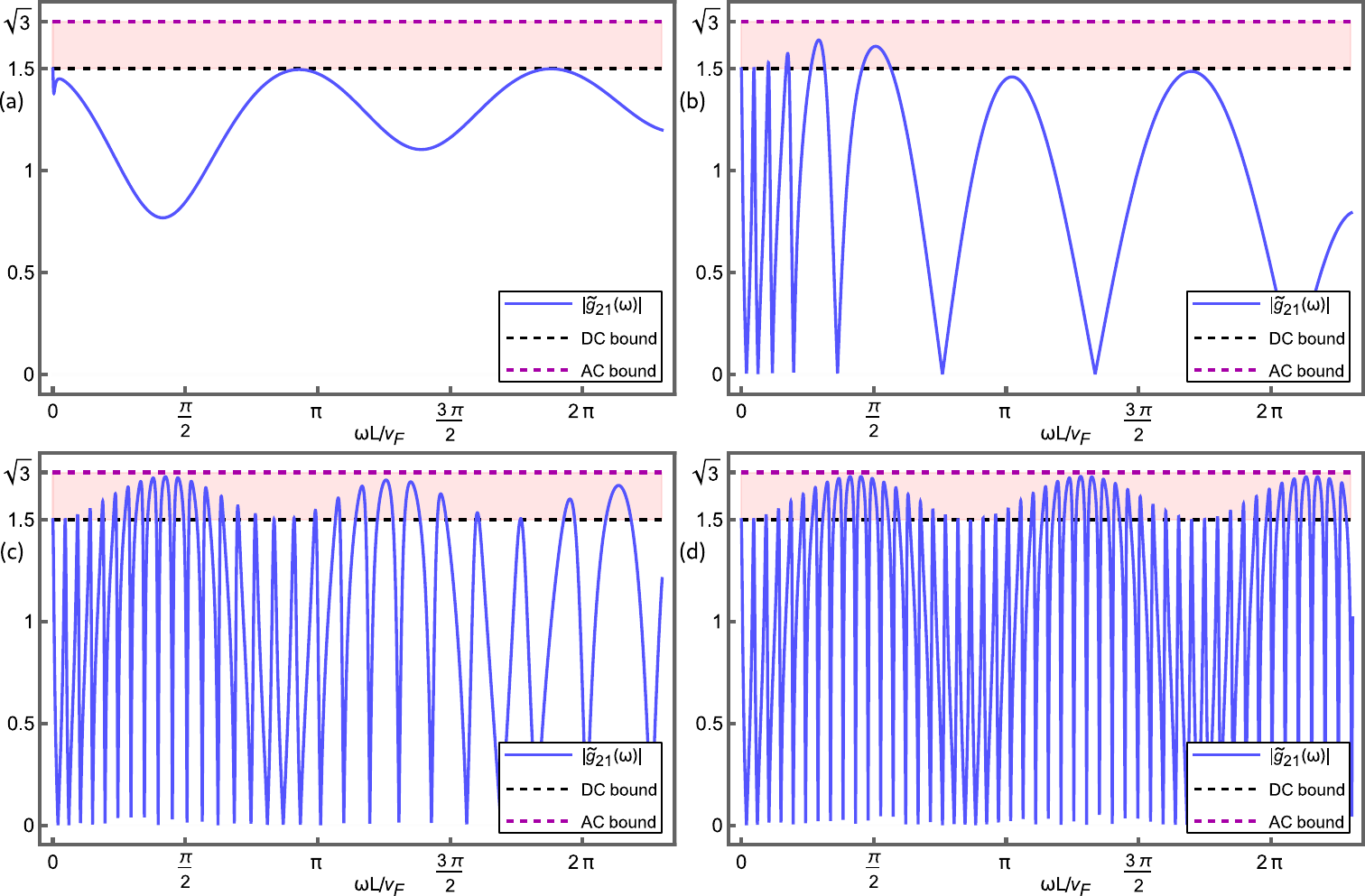}
    \caption{$|\Tilde{g}_{21}(\omega)|$ is plotted as a function of frequency $\omega$ for four values of $\eta$: (a) 0.1, (b) 10, (c) 100, and (d) 1000, with $L = 1$.}
    \label{fig:LargeL}
\end{figure}
For a screened Coulomb interaction of the form $e^{-\eta|x-x'|}/|x-x'|$, $1/\eta$ characterizes the range of the interaction, while $L$ represents the length of the interacting region. Fig.~\ref{fig:LargeL} shows a plot for various values of $\eta$ with $L=1$. As the range of the interaction ($1/\eta$) decreases, meaning the screening gates are closer and more effective at suppressing the long-range nature of the interaction, more resonance peaks start to appear within the same frequency window. Note that in the fully screened limit with $\eta$ = 1000 (see Fig.~\ref{fig:LargeL}(d)), we recover our result with point-like interaction.
\section{Wave packet dynamics:} In this section, we take the complementary approach by studying the spatio-temporal evolution of an electron wave packet being injected at $x>L$ along the incoming QH edge. The injected wave packet is considered to have a Gaussian distribution: $\rho_{iI}(x)= \text{exp}(-(x-b)^2/2\sigma^2)/(\sigma\sqrt{2\pi
})$, with a width of `$\sigma$' and `$b$' denoting the centre of the injection point. If the width of a single electron pulse is small compared to the length of the interacting region, i.e., $\sigma < L$, the electron wave packet fractionalizes into multiple pulses along the outgoing QH edges. 
\begin{figure}[h]
\includegraphics[width=0.98\textwidth]{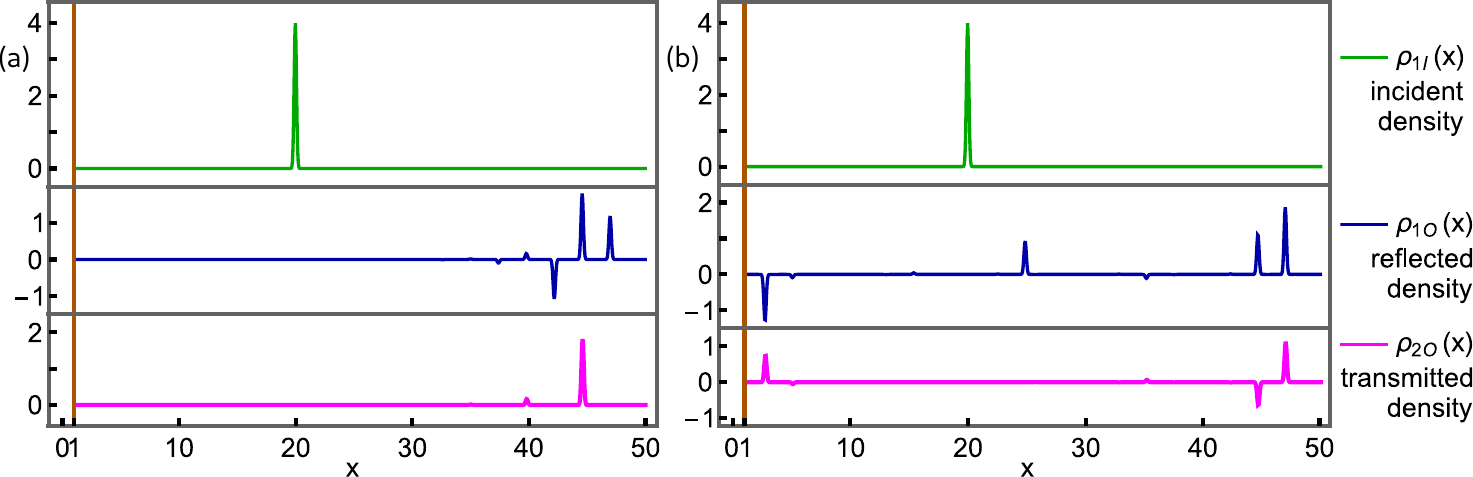}
\caption{Density profiles with an interacting region extending from $x=0$ to 1 and total length of the QH edges 100 for a $\nu_1 = 1 \text{ and } \nu_2=1/3$ QH junction and $v_F$ is taken to be one. The density profile is shown when the electron is injected from $\nu=1$ side, and with (a) only intraedge point-like interaction $\alpha = 0.55,\beta=\gamma=0$, (b) both intraedge and interedge interactions $\alpha = 0.55,\beta=0.14, \gamma=0.589$}
    \label{fig:pointwavepacket}
\end{figure}

Interestingly, in the presence of only intraedge point-like interaction ($\beta = \gamma =0$), no hole or negative pulse is reflected at the $x=L$ boundary in the transmitted edge channel (see fig.~\ref{fig:pointwavepacket}(a)). However, in the presence of interedge point-like interaction ($\beta,\gamma \neq 0$), both the transmitted and reflected edge channels acquire fractionalized negative pulses in response to a positive incoming pulse (see fig.~\ref{fig:pointwavepacket}(b)). These fractionalized negative pulses in the transmitted edge channel can be considered as a signature of the presence of interedge interaction in the system in the chiral time-resolved wave-packet measurement. In the long time limit, the density integral over the transmitted or reflected edge channel reduces to its DC value of the scattering matrix at $x=0$, as expected, which adds up to one, confirming current conservation.
\begin{figure}[h]
\includegraphics[width=0.98\textwidth]{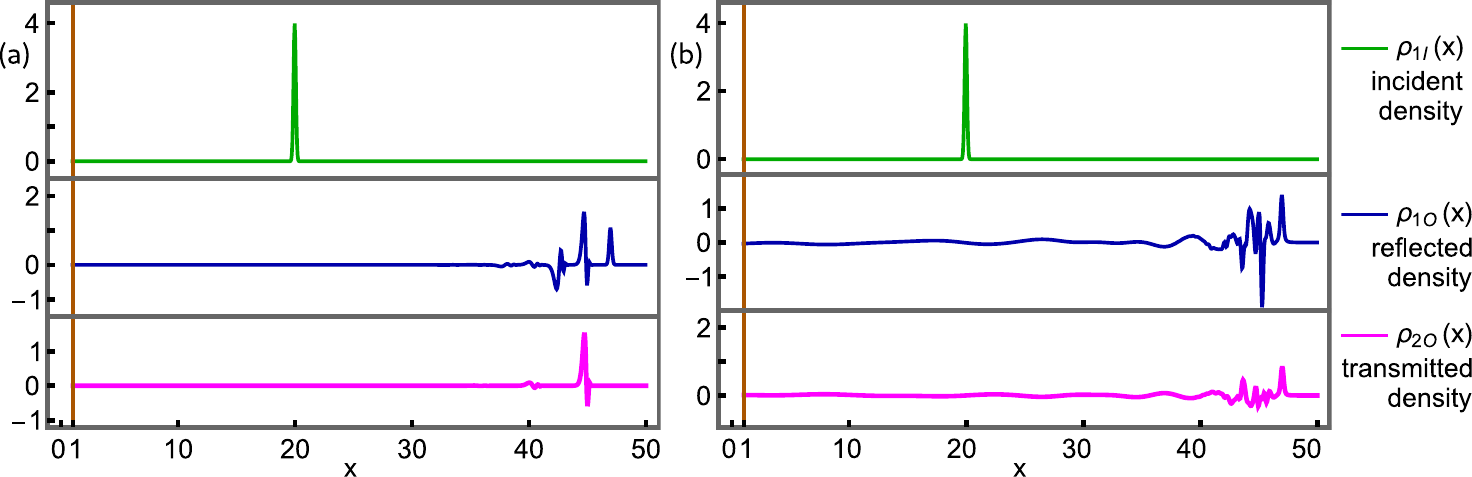}
\caption{Density profiles with an interacting region extending from $x=0$ to 1 and total length of the QH edges 100 for a $\nu_1 = 1 \text{ and } \nu_2=1/3$ QH junction and $v_F$ is taken to be one. The density profile is shown when the electron is injected from $\nu=1$ side, and with (a) only intraedge screened Coulomb interaction $\alpha = 0.55,\beta=\gamma=0$, (b) both intraedge and interedge screened Coulomb interactions $\alpha = 0.55,\beta=0.14, \gamma=0.589$}
    \label{fig:longwavepacket}
\end{figure}

The wave packet dynamics in the presence of intraedge (see fig.~\ref{fig:longwavepacket}(a)) along with interedge (see fig.~\ref{fig:longwavepacket}(b)) screened Coulomb interaction have been studied, where in both cases fractionalized negative pulses appear in the transmitted channel.}
\end{document}